\documentclass[10pt, pre, amsmath, onecolumn, showpacs, superscriptaddress]{revtex4-1}
\usepackage{graphicx}
\usepackage{hyperref}
\usepackage{amsfonts}
\usepackage{bm}
\usepackage{multirow}
\usepackage[rgb]{xcolor}
\usepackage{tikz}
\usetikzlibrary{shapes,shapes.multipart}
\allowdisplaybreaks
\DeclareMathAlphabet{\mathpzc}{OT1}{pzc}{m}{it}
\usepackage[mathscr]{eucal}

\begin{document}
\title{ Long range correlations in stochastic transport with energy and momentum conservation}
\author{Anupam Kundu$^{1,2}$, Ori Hirschberg$^{1,3}$, David Mukamel }
\affiliation{Department of Physics of Complex Systems, Weizmann
Institute of Science, 76100 Rehovot, Israel\\
$^2$International center for theoretical sciences, TIFR, Bangalore - 560012, India\\
$^3$Department of Physics, Technion, 3200003 Haifa, Israel 
}

\date{\today}
\begin{abstract}

We consider a simple one dimensional stochastic model of heat transport which locally conserves
both energy and momentum and which is coupled to heat reservoirs with different temperatures at 
its two ends. The steady state is analyzed and the model is found to obey the Fourier law with finite 
heat conductivity. In the infinite length limit, the steady state is described locally by an 
equilibrium Gibbs state. However finite size corrections to this local equilibrium state are present. We analyze
these finite size corrections by calculating the on-site fluctuations of the momentum and the two point correlation
of the momentum and energy. These correlations are long ranged and have scaling forms which are computed explicitly.
We also introduce a multi-lane variant of the model in which correlations vanish in the steady state. 
The deviation from local equilibrium in this model as expressed in terms of the on-site momentum fluctuations is calculated in the large length limit.
\end{abstract}

\maketitle


\noindent
\section{Introduction}
\noindent
More than two centuries after its initial formulation, Fourier’s law still poses major theoretical challenges. The law stipulates that the energy current through 
a material is linearly proportional to the temperature gradient maintained at the boundaries. While widely successful on a phenomenological level, explaining how 
this macroscopic law arises from complex microscopic interactions is generally an open problem, and one that has been a subject of great interest for the last 
few decades (reviewed, e.g., in \cite{AD08, Lepri03} ). 
Starting from a microscopic description, a successful approach for understanding fluctuations in such situations
would ideally require analyzing and possibly computing explicitly the joint probability distribution of all the relevant degrees of freedom present
in the system, \emph{i.e.} fully characterizing the non-equilibrium state. In practice this is a very hard task to perform with present
day mathematical tools.

Due to the lack of a general framework for characterizing non-equilibrium systems,
studies of simple models provide great insight into their nature. There are only a few model systems
for which one can perform this task successfully.  One such model is the Lebowitz-Reider-Leib model \cite{Rieder67}
where the energy transport through a harmonic chain connected to stochastic reservoirs at its two ends has been considered. 
It has been shown that the stationary invariant measure in phase space is a
multivariate Gaussian and due to the lack of collisions there is no diffusive transport of energy. As a result the phenomenological Fourier law
which governs the heat current is not satisfied and anomalous energy transport takes place. Extensions of this model
have been studied whereby anharmonicity,  disorder or both have been introduced to provide a source of collisions among the harmonic degrees of freedom.
In some of the cases studied, these collisions do not lead to diffusive transport resulting in anomalous
behavior in low dimensions \cite{AD08,AC10,Saito10,GG08,Lepri97}. For general Hamiltonian systems, explicating Fourier's law has not been achieved completely.
In fact, studying non-equilibrium states even for simple deterministic non-linear systems is still a theoretical challenge \cite{Bonetto}.

As a complementary approach, simplifications have been traditionally made by introducing stochasticity into the microscopic dynamics of the system.
Various lattice gas models provide simple examples for such stochastic models of transport. In these models, the microscopic dynamics
is generally described in terms of dynamical processes with specified rates that satisfy some general requirements like local detailed balance,
finite range of interaction and translational invariance \cite{Eyink90}. Such stochastic models are assumed to provide a
reduced (microscopic) description of the intricate chaotic aspects of the microscopic deterministic dynamics. Although
stochastic models are in principle simpler to analyze
as compared with deterministic dynamical models, even here there are few such models for which one can provide a complete description of the
non-equilibrium steady state. One such example is the zero range process, in which the dynamics of mass leaving a site depends only on the occupation
number at that site. For this process the steady state distribution is known to factorize under certain conditions, into the product of single-site distributions 
and thus many analytical results can be obtained \cite{Evans04}. However for more complicated models this factorization property no longer exists.

An interesting class of models in which transport properties have been studied is the Kipnis-Marchioro-Presutti (KMP) model \cite{KMP82}. 
This model is composed of a $1d$ lattice of length $L$ where each site $i$ carries an energy variable $\epsilon_i$. The model evolves by a random sequential updates whereby a pair of nearest
neighbor sites is chosen at random and the energies of the two sites are mixed keeping the total energy fixed. The lattice is coupled at the left and right
ends to two reservoirs with temperatures $T_l$ and $T_r$, respectively. In the steady state the model has been shown to obey the Fourier law and the
long range nature of the energy-energy correlation function has been demonstrated. The model has a single (bulk) conserved quantity, the energy. This fact plays a crucial
role in determining the nature of its steady state. Similar stochastic dynamical processes have been incorporated in harmonic chain models keeping both
the momentum and energy variables conserved \cite{Basile06, Bernardin08, Basile07,Basile09}. In these models
the mixed nature of the dynamics, composed of both deterministic Hamiltonian moves, and stochastic processes, yields a steady state with anomalous energy transport. 
It is of great interest to explore the effect of more than one conserved quantity on the steady state of systems which evolve only under stochastic dynamics, and 
study the transport properties and the long range nature of the correlation functions.

In this paper we consider such a model, with purely stochastic dynamics of energy transport across a one dimensional lattice of size $L$ and two conservation laws.
Each lattice site carries a ``momentum'' variable $p$, which, according to the stochastic dynamics, is mixed with the momenta of the two nearest
neighbor sites with some rates, keeping the total momentum and kinetic energy fixed. The lattice is connected at its left and right ends to stochastic reservoirs
with temperatures $T_l$ and $T_r$ respectively.
For this system, we study spatial long-range correlations of momenta and energy in the non-equilibrium steady state.
When the temperatures of both
the reservoirs are equal, the system reaches an equilibrium state in which the the joint distribution of the momenta of different sites is given by
a product of Gaussian distributions for each site. When the temperatures of the two reservoirs are different, on thew other hand, the
system reaches a nonequilibrium steady state (NESS) where the joint distribution can no longer be factorized. However, for large systems ($L \to \infty$) with a finite 
arbitrary temperature difference,
the system locally reaches an equilibrium state \emph{i.e.} the marginal distribution of the momentum of the $i$th site (in the bulk) can be approximately
described by a Gaussian distribution with a local temperature. Such states are called local equilibrium (LE) states. For systems with
finite but large $L$, one observes deviation from the local Gaussian distributions and this deviation may be quantified by the
correlations between different sites. In this paper we compute such deviations explicitly.

The paper is organized as follows. In sec. \ref{Model} we first introduce the model and the dynamics. After that we give a summary of our results. 
The average temperature profile is calculated in sec. \ref{e-profile-sec}. Next we compute correlations in sec. \ref{F-S-correc} as finite size corrections. 
We first consider a simpler multi-channel version of our original model in sec. \ref{MF}. In this model, when the channel number becomes large, the correlations
between momenta of different sites become negligible. This is however not true in the original model, where momenta of different particles are correlated. We compute 
these correlations in sec. \ref{corr-only-e-drive}. Finally, sec. \ref{conclu} concludes the paper.

\section{Stochastic model and summary of the results}
\label{Model}
\begin{figure}[h]
\includegraphics[scale=0.35]{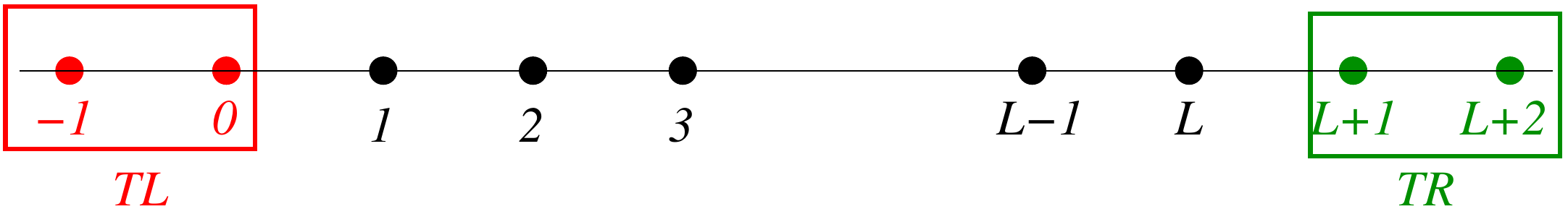}
\caption{ (Color Online) Schematic diagram of the model}
\label{fig1}
\end{figure}

\noindent
We consider a lattice having $L+4$ sites $(-1,0,1,...,L,L+1,L+2)$, where each site, $i$, carries some momenta $p_i$.
These momenta can change only through a three particle ``collision'' which conserves both total momentum and energy of the
three chosen sites.
The two edge sites at the left boundary and the two edge sites at the right boundary are considered as bath sites with temperatures $T_l$ and $T_r$ respectively.
A schematic cartoon of the system is given in fig. {\ref{fig1}}.

\noindent
In detail the dynamics is the following :
\begin{itemize}
 \item (a) First a site from the set $[0,1,2,...,L,L+1]$ is chosen at random, say site $i$. Let the momentum of the site be
           $p_i$ and the energy be $\frac{p^2_i}{2}$.
           
 \item (b) Consider its two nearest neighbor sites $i-1$ and $i+1$:  The momenta of these
           three sites $\{p_{i-1},p_i,p_{i+1}\}$ are now randomly mixed in such a way that both the total momentum $P_i=p_{i-1}+p_i+p_{i+1}$ and
           the total energy $2E_i=p_{i-1}^2+p_i^2+p_{i+1}^2$ remain conserved. This implies that the resulting momenta $\{q_{i-1},q_i,q_{i+1}\}$ after 
           the mixing process should satisfy
           \begin{eqnarray}
            & &q_{i-1}+q_i+q_{i+1}=P_i,\label{momenta} \\
            \text{and} \quad & &\frac{q^2_{i-1}}{2}+\frac{q^2_{i}}{2}+\frac{q^2_{i+1}}{2}=E_i. \label{energy}
           \end{eqnarray}
           Equations (\ref{momenta}) and (\ref{energy}) define a plane and a sphere in the 3-dimensional q-space, whose intersection is a circle.
           Any random vector $\vec{q}\equiv (q,q',q'')$ with its tip in this circle would produce the desired mixing. The intersection circle
           can be parametrized, with the help of little algebra, as follows:
           \begin{eqnarray}
            q_{i-1}&=&\frac{1}{3}[P_i+R_i \cos \theta] \nonumber \\
            q_i&=&\frac{1}{3}[ P_i - R_i(\cos \theta - \sqrt{3}\sin \theta)/2] \label{rules} \\
            q_{i+1}&=&\frac{1}{3}[ P_i - R_i(\cos \theta + \sqrt{3}\sin \theta)/2], \nonumber
           \end{eqnarray}
            where $R=\sqrt{12E_i-2P_i^2}$ and $\theta \in [0,2\pi)$. We focus on dynamics where $\theta$ is chosen uniformly, corresponding to a
            uniformly chosen point on the circle.
            
\item  (c) The above rule (b) apply for any site except for $i=1,2,L-1$ and $L$. ``Collisions'' at these
           sites involve $i=-1,0,L+1$ and $L+2$, which represent the heat baths. 
           To include the effect of the heat baths, whenever sites $i=-1,0$ appear among the three chosen sites, their momentum is first chosen 
           from the distribution
           \begin{equation}
            g_l(p)=\frac{1}{\sqrt{2\pi T_l}}e^{-p^2/2T_l},~~\text{for}~~i=-1,0. \label{ht-bth-l}
           \end{equation}
           Similarly, when the sites $i=L+1$ and $L+2$ appear among the three chosen sites, their momenta is first chosen from the distribution 
           \begin{equation}
            g_r(p)=\frac{1}{\sqrt{2\pi T_r}}e^{-p^2/2T_r},~~\text{for}~~i=L+1,L+2. \label{ht-bth-r}
           \end{equation}
           After these random values are chosen, the momenta of the three chosen sites are mixed following rule (\ref{rules}).
\end{itemize}
We  note that the dynamics discussed above conserves both momentum and energy through a random mixing of the momenta variables.
Similar random mixing of energy or momentum have been previously considered in other contexts. For example, in the KMP model \cite{KMP82}
energies of the neighboring sites redistribute
themselves randomly in such a way that the total  energy is conserved. Other variants of this model, known in general as conserved mass models
\cite{Rajesh00}, have also been considered. Models with stochastic interchange added to an existing Hamiltonian dynamics was considered in the
context of anomalous energy transport through coupled oscillators \cite{Basile06, Basile07, Basile09, Lepri09, Lepri10}.
In most of these cases, the ``collision'' process between different sites
involve {\it two} neighboring sites. In our model we consider ``collisions'' among {\it three} neighboring sites. In the context of transport
such three particle ``interaction'' models were considered by S. K. Ma in \cite{Ma83}, where he studied energy
transport starting from Boltzmann equation and considered three particle collision term as a mechanism of thermalization.
Basile et al. \cite{Basile06} have added such a three particle stochastic ''collision'' term to an existing Hamiltonian part while looking at
energy transport through oscillator chains. A similar ``three particle collision''
model was considered by Lubini \emph{et al} \cite{Lubini14} in the context of coarsening dynamics.
Recently models with two conserved quantities have been examined to study condensation of large fluctuations in linear statistics \cite{Nossan}.

In the rest of the paper we present our analysis of the statistics of the momenta corresponding to the above mentioned dynamics.
The summary of our results is the following :
\begin{itemize}
 \item (i) For our model with both the energy and momentum conservation (except at the boundary), 
we obtain the following linear behavior of the local energy $e_i=\langle p_i^2 \rangle$ in the NESS:
\begin{equation}
 e_{i}= T + \frac{\Delta T}{2} \left(1-2\frac{i}{L} \right),~~i=1,2,...,L,~~~\text{with}~~ T=\frac{T_l+T_r}{2},~~\Delta T= T_l-T_r, \nonumber
\end{equation}
for large $L$. Note that the variable $e_i$ is twice the actual average energy $\langle p_i^2 \rangle/2$. Throughout this paper we will consider $e_i$ as energy variable.
For finite $L$, however, there are deviations from the above linear  behavior. 
We also study this deviation by computing an exact expression for the energy profile for arbitrary system length $L$.
The corresponding energy current in the NESS decays as $\sim 1/L$ for large $L$
indicating ''Fourier'' behavior. We find that the diffusivity is constant, \emph{i.e} independent of the local energy.

\item (iii) In the steady state of this model, there are correlations among distant sites which provide a measure of the deviation from LE. 
We find that such correlations decay as $\sim 1/L$ and at this order
they are long ranged. More precisely, we find that the energy-energy correlation
$C_{i,j}=\langle p_i^2 p_j^2 \rangle- \langle p_i^2\rangle\langle p_j^2\rangle$ and the kurtosis
$K_i = \langle p_i^4\rangle - 3\langle p_i^2 \rangle^2$
have the following scaling forms
\begin{equation}
 K_i = \frac{1}{L} \mathscr{K} \left(\frac{i}{L} \right ),~~C_{i,j} = \frac{1}{L} \mathscr{C}\left(\frac{i}{L},\frac{j}{L} \right), \nonumber
\end{equation}
for large $L$. Taking an appropriate continuum limit of the discrete equations satisfied by $C_{i,j}$
and $K_i$, we find that the scaling function $\mathscr{C}(x,y)$ satisfies a differential equation which can be interpreted as a
problem of finding the electrostatic potential created by a uniform charge density along the diagonal of a square of unit length. Finally, we find the 
explicit expressions of the scaling functions $\mathscr{K}(x)$ and $\mathscr{C}(x,y)$ for $(x,y)\in[0,1]^2$ as :
\begin{eqnarray}
\mathscr{C}(x,y) &=& 2~\Delta T^2~\begin{cases}
                                                 & ~x(1-y),~~~x>y \\
                                                 & ~y(1-x),~~~y>x
                                                \end{cases},  \nonumber \\
\mathscr{K}(x)&=& 6~\Delta T^2~x(1-x).
\end{eqnarray}

 \item (ii) We also study a multi-lane variant of the original single lane model in which all correlations between different sites vanish.  We find that the 
 deviations from local equilibrium in this case for large but finite $L$ exist. As a measure of these deviations the kurtosis,  
$K_i = \langle p_i^4\rangle - 3\langle p_i^2 \rangle^2$, is calculated and found to decay as $\sim 1/ L^2$ for large $L$. 

\end{itemize}

\subsection{Master equation and NESS}
\label{pdfNESS}
\noindent
From the dynamics discussed in the previous section one can write the following master equation satisfied by
the joint distribution $P^{(L)}(\{p_i\}, t)$ of the momenta of the $L$ bulk sites as follows :
Let $P^{(L)}(\{p_i\}, t+dt)$ be the joint probability of configuration $\{p_i\}\equiv (p_1,p_2,...,p_L)$ at time $t+dt$, then
\begin{eqnarray}
&&P^{(L)}(\{p_i\}, t+dt) =  (1-(L+2)dt)~\times~P^{(L)}(\{p_i\}, t) + \sum_{\{q_i\}} ~ \mathcal{K}({\bf p}|{\bf q})dt~\times~P^{(L)}(\{q_i\}, t),
\end{eqnarray}
where the first term on the right hand side corresponds to probability that the configuration $\{p_i\}$ does not change while the second term corresponds 
to the probability of a
transition from some configuration $\{q_i\}$ to $\{p_i\}$ during the time interval $dt$.
There are $(L+2)$ different processes in which the system can reach the configuration $\{p_i\}$ at time $t+dt$, among which there are $(L-2)$ bulk
processes that occur with ``collision'' rate $\mathcal{K}({\bf p}|{\bf q})$. The remaining four processes represent interaction with baths at the boundaries.
Taking $dt \to 0$ on both sides and using the following notations
\begin{eqnarray}
&&{\bf q}_{i}\equiv(q_{i-1},q_i,q_{i+1})~~d{\bf q}_{i}\equiv dq_{i-1}~dq_{i}~dq_{i+1},~P_i=p_{i-1}+p_i+p_{i+1}, \nonumber \\
&&Q_i=q_{i-1}+q_i+q_{i+1},~E_i^p=\frac{p_{i-1}^2+p_i^2+p_{i+1}^2}{2}~~\text{and}~~
E_i^q=\frac{q_{i-1}^2+q_i^2+q_{i+1}^2}{2}. \label{notations1}
\end{eqnarray}
we arrive at
\begin{equation}
\left (\frac{\partial }{\partial t} + (L+2) \right) P^{(L)}(\{p_i\}, t)= ~\hat{\mathcal{R}}P^{(L)}(\{q_i\},t),
\end{equation}
where,
\begin{eqnarray}
&&\hat{\mathcal{R}}P^{(L)}(\{q_i\},t)= \sum_{i=2}^{L-1} \int d {\bf q}_i ~
\mathcal{K}_i({\bf p}_i | {\bf q}_i)~P^{(L)}(\{p_1,p_2,...,q_{i-1},q_i,q_{i+1},..,p_{L}\}, t) \nonumber \\
&&~~~~~~~~ +~\int dp_{-1} \int dp_{0}\int d {\bf q}_0 ~
\mathcal{K}_0({\bf p}_0 | {\bf q}_0)~g_l(q_{-1})g_l(q_0)P^{(L)}(\{q_1,p_2,...,p_{L}\}, t) \nonumber \\
&&~~~~~~~~ +~\int dp_{0}\int d {\bf q}_1 ~
\mathcal{K}_1({\bf p}_1 | {\bf q}_1)~g_l(q_0)P^{(L)}(\{q_1,q_2,p_3,...,p_{L}\}, t)\label{master-eq} \\
&&~~~~~~~~ +~\int dp_{L+1}\int d {\bf q}_{L} ~
\mathcal{K}_L({\bf p}_L | {\bf q}_L)~P^{(L)}(\{p_1,...,p_{L-2},q_{L-1},q_{L}\}, t)g_r(q_{L+1})\nonumber \\
&&~~~~~~~~ +~\int dp_{L+2} \int dp_{L+1}\int d {\bf q}_{L+1} ~
\mathcal{K}_{L+1}({\bf p}_{L+1} | {\bf q}_{L+1})~P^{(L)}(\{p_1,...,p_{L-1},q_{L}\}, t)g_r(q_{L+1})g_r(q_{L+2}), \nonumber  \\
&& \text{with}  \nonumber \\
&&~~~~~~~~~~~~~~\mathcal{K}_i({\bf p}_i | {\bf q}_i)\equiv
\frac{\sqrt{3}}{2 \pi}\delta(P_i-Q_i)~\delta(E^p_{i}-E^q_i). \label{notations}
\end{eqnarray}
The distributions $g_l$ and $g_r$ are given in equations (\ref{ht-bth-l}) and (\ref{ht-bth-r}) respectively. The first term on the right hand side of the above
equation represents the mixing in the bulk $2 \leq i \leq L-1$ (see sec. \ref{Model}), whereas the $2$nd and $3$rd terms correspond to the interaction
with the left heat bath and the $4$th and $5$th terms correspond to the interaction with the right heat bath.
Note that $P_i$ and $E_i^p$ in eq. (\ref{notations1}) represent the total momenta and total energy of the three chosen sites, respectively,
before collision and $Q_i$ and $E_i^q$ represent the total momenta and total energy, respectively, after collision.
The expression of the collision kernel $\mathcal{K}_i({\bf p}_i | {\bf q}_i)$ in eq. (\ref{notations}) explicitly shows
momentum and energy conservation at each collision. The factor $\frac{\sqrt{3}}{2\pi}$ is chosen so as to normalize the distribution
\emph{i.e.} $\prod \int dp_i~P^{(L)}(\{p_i\},t) = 1$.

The master equation in the steady state is obtained from $\frac{\partial }{\partial t}P^{(L)}(\{p_i\})=0$ as
\begin{equation}
 (L+2)~P^{(L)}(\{p_i\})= ~\hat{\mathcal{R}}P^{(L)}(\{q_i\}). \label{SS-mstr-eq}
\end{equation}
Integrating over all $p_j$ except $p_i$ on both sides of this equation we obtain
the following equation satisfied by the marginal distribution $P^{(1)}(p_i) = \prod_{j\neq i}\int dp_j~P^{(L)}(\{p_j\})$ in the NESS:
\begin{eqnarray}
 &&P^{(1)}(p_i)=\frac{1}{3}~
\Big{[} \int dp_{i-2} \int dp_{i-1} ~\int d {\bf q}_{i-1}~\mathcal{K}_{i-1}({\bf p}_{i-1} | {\bf q}_{i-1})~P^{(3)}({\bf q}_{i-1})
+ \int dp_{i-1} \int dp_{i+1} ~\int d {\bf q}_{i}~\mathcal{K}_i({\bf p}_i | {\bf q}_i)~P^{(3)}({\bf q}_{i}) \nonumber  \\
&&~~~~~~~~~~~~~~~~~~~~~~~~~~~~~~~~~~~~~~~~~+\int dp_{i+1} \int dp_{i+2}~\int d {\bf q}_{i+1}
~\mathcal{K}_{i+1}({\bf p}_{i+1} | {\bf q}_{i+1})~P^{(3)}({\bf q}_{i+1}) \Big{]},
\label{NESSpdf}
\end{eqnarray}
where $P^{(3)}({\bf q}_{i}) \equiv P^{(3)}(q_{i-1},q_i,q_{i+1})$ is 
the marginal distribution obtained by integrating $P^{(L)}(\{q_j\})$ over all $q_j$'s except $q_{i-1},q_i$ and $q_{i+1}$.

\noindent
{\bf Thermodynamic limit}: Solving the above equation for $P^{(1)}(p)$ is difficult in general. However one can check that in $L \to \infty$ limit,
the following function
\begin{equation}
P^{(1)}(p_i)\simeq g_i(p_i,T_i)=\frac{1}{\sqrt{2\pi T_i}}~e^{-\frac{p^2}{2T_i}},
\end{equation}
for the marginal distribution of the momentum of $i$th site, with
\begin{equation}
T_i= T_l - (T_l-T_r) \frac{i}{L}, \label{profiles}
\end{equation}
satisfies the above equation (\ref{NESSpdf}) to leading order in $1/L$. This suggests that locally we have
equilibrium with temperature $T_i$. In the next section we show that indeed in the large $L$ limit the temperature profile is given by (\ref{profiles}).
In fact, to leading order in $1/L$, the full joint distribution of all the
momenta can be given by $P^{(L)}(\{p_i\}) \simeq \prod_i g(p_i)$.
Corrections to this local equilibrium measure may appear from the correlations among the momenta in finite
size systems, and these corrections decrease to zero with increasing system size $L$. In the rest of the paper we study 
these correlations in more detail. Before doing so, let us first look at the average
local energy and the steady state current.

\section{Energy profile and current in the NESS}
\label{e-profile-sec}
\begin{figure}{}
\includegraphics[scale=0.35]{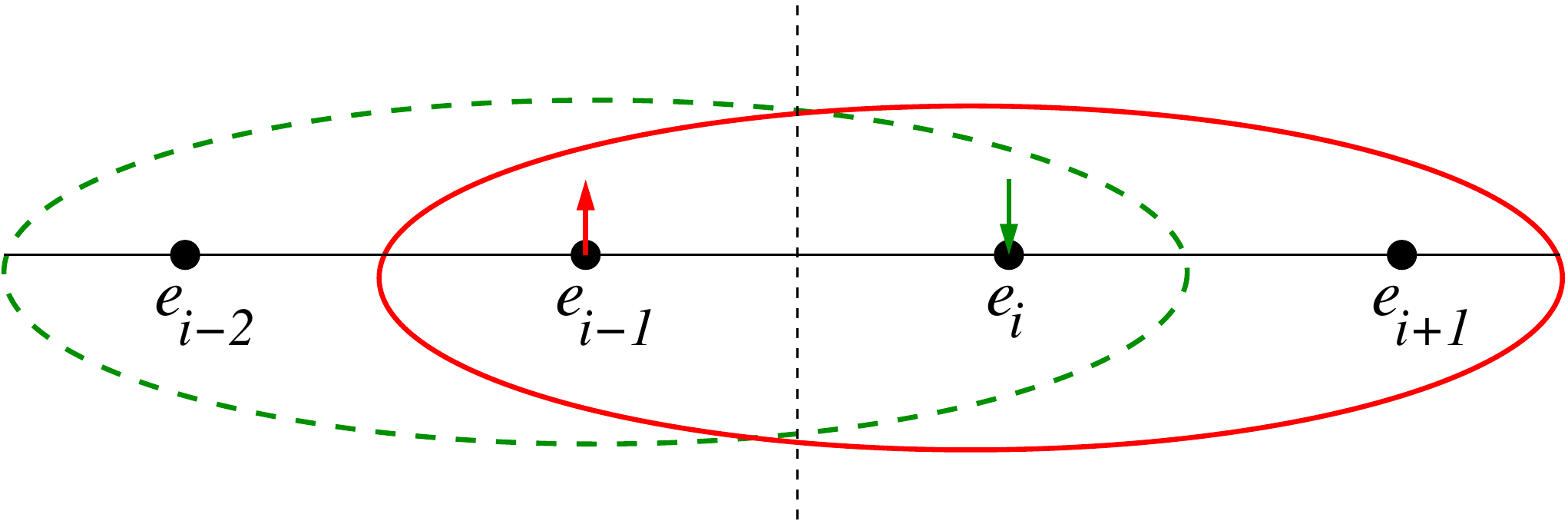}
\caption{ (Color Online) Contribution to the current $J_i^{(e)}$ due to the mixing of energies. The green arrow represents current entering
site $i$ in the interaction among $(i-2,~i-1,~i)$ (green dashed ellipse) whereas the red arrow represents the current leaving site $(i-1)$
in the interaction among $(i-1,~i,~i+1)$ (red solid line ellipse).}
\label{fig2}
\end{figure}

\noindent
Given the distribution $P^{(1)}(p_i)$ in NESS, the average energy $e_i=\langle p_i^2 \rangle$ can be computed from
$e_i=\int_{-\infty}^{\infty}~dp_i~p_i^2~P^{(1)}(p_i)$ and the marginal distribution $P^{(1)}(p_i)$ can in principle be obtained
by solving eq. (\ref{NESSpdf}). Instead, using the structure of eq. (\ref{NESSpdf}),
one can obtain a recursion relation of $e_i$. To get that, we multiply both sides of eq. (\ref{NESSpdf}) by $p_i^2$ and integrate
over $p_i$ from $-\infty$ to $\infty$ yielding
\begin{eqnarray}
 3e_i&=&\int_{-\infty}^{\infty}~dp_i~p_i^2~ \Big{[} \int dp_{i-2} \int dp_{i-1} ~\int d {\bf q}_{i-1}
~\mathcal{K}_{i-1}({\bf p}_{i-1} | {\bf q}_{i-1})~P^{(3)}({\bf q}_{i-1})
+ \int dp_{i-1} \int dp_{i+1} ~\int d {\bf q}_{i}~\mathcal{K}_i({\bf p}_i | {\bf q}_i)~P^{(3)}({\bf q}_{i}) \nonumber  \\
&&~~~~~~~~~~~~~~~~~~~~~~~~~~~~~~~~~~~~~~~~~+\int dp_{i+1} \int dp_{i+2}~\int d {\bf q}_{i+1}
~\mathcal{K}_{i+1}({\bf p}_{i+1} | {\bf q}_{i+1})~P^{(3)}({\bf q}_{i+1}) \Big{]}. \label{e-i-1}
\end{eqnarray}
We now have to evaluate the three integrals, denoted by $\mathcal{I}_{i-1},~\mathcal{I}_{i}$ and $\mathcal{I}_{i+1}$, respectively, in the right hand side of the
above equation. Let us consider the middle integral
\begin{equation}
 \mathcal{I}_i= \int_{-\infty}^{\infty}~d{\bf p}_i~\int_{-\infty}^{\infty} d {\bf q}_{i}~p_i^2~ \mathcal{K}_i({\bf p}_i | {\bf q}_i)~P^{(3)}({\bf q}_{i})
=\frac{\sqrt{3}}{2\pi}\int_{-\infty}^{\infty}~d{\bf p}_i~\int_{-\infty}^{\infty} d {\bf q}_{i}~p_i^2~
~\delta(P_i-Q_i)~\delta(E^p_{i}-E^q_i)~P^{(3)}({\bf q}_{i}), \label{mcalI0}
\end{equation}
where the notations in eq. (\ref{notations1}) are used. We first perform the integrations over $(p_{i-1}, p_i, p_{i+1})$, \emph{i.e}
$\int_{-\infty}^{\infty}~d{\bf p}_i~p_i^2~ \mathcal{K}_i({\bf p}_i | {\bf q}_i)$. To carry out this integration, we use the following transformation
\begin{eqnarray}
            p_{i-1}&=&\frac{1}{3}[P_i+R_i^p \cos \theta_i^p] \nonumber \\
            p_i&=&\frac{1}{3}[ P_i - R_i^p(\cos \theta_i^p - \sqrt{3}\sin \theta_i^p)/2] \label{transformation} \\
            p_{i+1}&=&\frac{1}{3}[ P_i - R_i^p(\cos \theta_i^p + \sqrt{3}\sin \theta_i^p)/2], \nonumber
           \end{eqnarray}
with $R_i^p=\sqrt{12E_i^p-2P_i^2}$. One can easily check that the Jacobian of the above transformation is $\frac{1}{6\sqrt{3}}$ \emph{i.e.}
\begin{equation}
 dp_{i-1}dp_idp_{i+1}=\frac{1}{6\sqrt{3}} R_i^pdR_i^p dP_i d\theta_i^p. \label{Jaco}
\end{equation}
This yields
\begin{eqnarray}
\mathcal{I}_i &=& \frac{1}{2\pi}\int_{-\infty}^{\infty} d {\bf q}_{i}~\int_0^{2\pi}d\theta_i^p~\frac{1}{9}
~\left[ Q_i - R_i^q(\cos \theta_i^p - \sqrt{3}\sin \theta_i^p)/2\right]^2~P^{(3)}({\bf q}_{i}) \nonumber \\
&=&\frac{1}{3}\left( \langle q_{i-1}^2\rangle + \langle q_{i}^2\rangle + \langle q_{i+1}^2\rangle\right)
= \frac{1}{3}(e_{i-1}+e_i+e_{i+1}), \label{mcalI}
\end{eqnarray}
where $R_i^q=\sqrt{12E_i^q-2Q_i^2}$ and $E_i^q$ and $Q_i$ are given explicitly in eq. (\ref{notations1}). Performing integrations over the 
$p$ variables before integrations over the $q$ variables in the other two integrals in eq. (\ref{e-i-1})
yields similar expressions for $\mathcal{I}_{i-1}$ and $\mathcal{I}_{i+1}$.
Adding them up we arrive
 at the following recursion relation among the average energies of different sites :
\begin{equation}
 6e_i=e_{i-2}+2e_{i-1}+2e_{i+1}+e_{i+2},~~i=1,2,...,L-1,L;~~e_{-1}=e_0=T_l,~~e_{L+1}=e_{L+2}=T_r. \label{e-prof-recur}
\end{equation}
This recursion relation can also be obtained directly from the energy conservation obeyed by the dynamics, as follows:
After each collision the total energy of the three sites becomes equally partitioned on average.
This can also be seen from eq. (\ref{mcalI}). As a result the average energy $J_i^{(e)}$ that crosses the bond $(i-1,~i)$
per unit time in the steady state (see fig. \ref{fig2}) is given by
\begin{eqnarray}
 J_i^{(e)}&=& \left(\frac{e_{i-2}+e_{i-1}+e_i}{3} -e_i \right) + \left(e_{i-1}-\frac{e_{i-1}+e_{i}+e_{i+1}}{3} \right)
= \frac{1}{3}(e_{i-2} + 3e_{i-1} -3e_i - e_{i+1}), \label{current_i_2}
\end{eqnarray}
 where the $e_i$'s are the steady state energies of sites $i=1,2,...,L$ and $e_{-1}=e_0=T_l$, $e_{L+1}=e_{L+2}=T_r$. The first term in eq. (\ref{current_i_2})
represents the current entering site $i$ resulting from a collision among $(i-2,~i-1,~i)$ whereas the second term
represents the current leaving site $(i-1)$ in the interaction among $(i-1,~i,~i+1)$.
In the steady state the current at any bond is independent of $i$, \emph{i.e.} $J=J_i^{(e)}=J_{i+1}^{(e)}$, which readily gives the recursion relation in
eq. (\ref{e-prof-recur}).
\begin{figure}[t]
\includegraphics[scale=0.3]{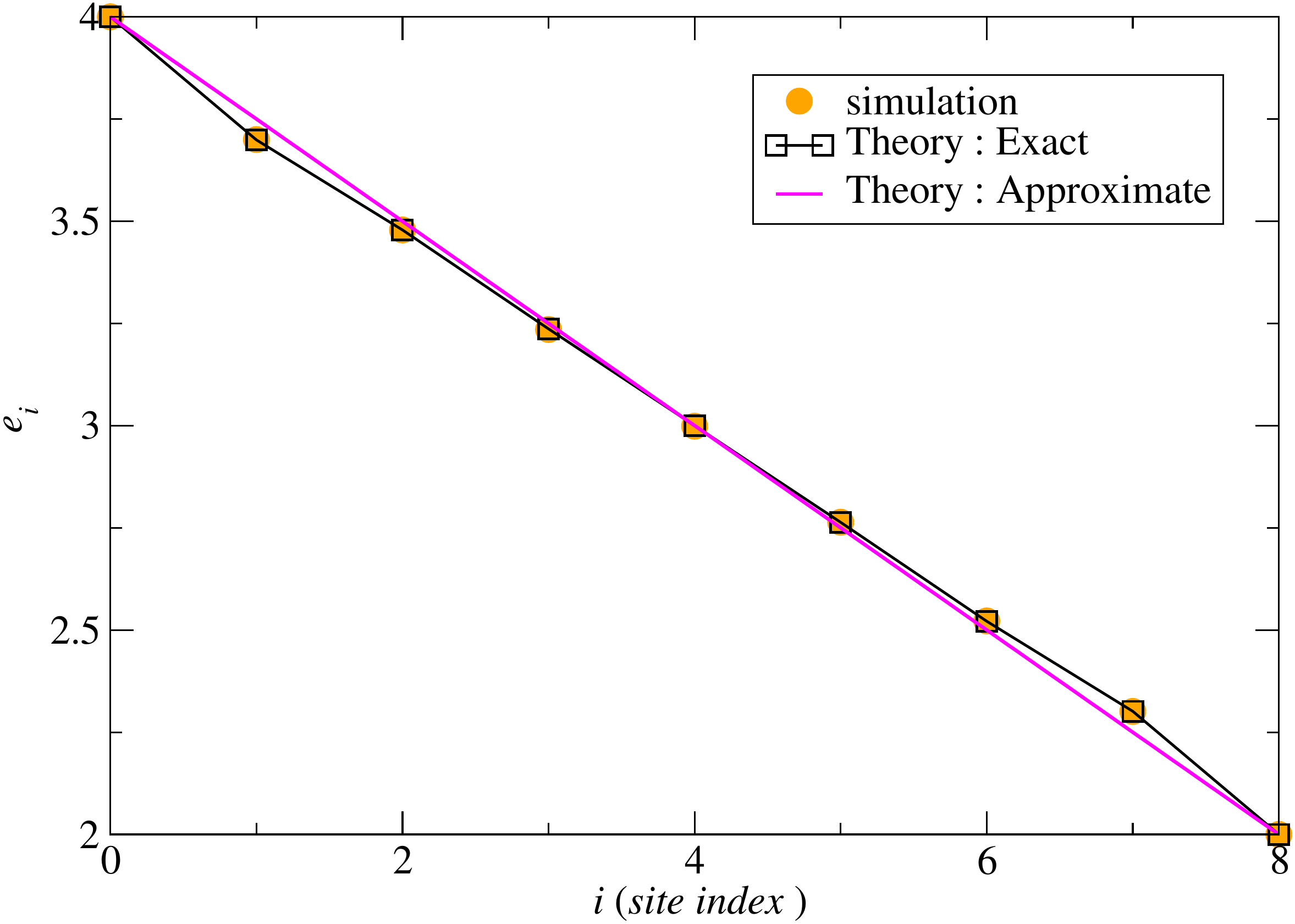}
\caption{ (Color Online) Comparison of the average energy profile obtained from numerical simulation and the analytical expression in eq. (\ref{e-profile})
for $L=7$ and $T_l=4,~T_r=2$. Note the deviations from the approximate linear profile at the boundaries.}
\label{fig3}
\end{figure}

Let us first solve this recursion relation in the $L \to \infty$ limit.
Assuming that for large $L$ the energy profile has the following scaling form $e_i=\mathscr{E}(i/L)$ and
inserting this form in eq. (\ref{e-prof-recur}) we get $\frac{d^2 \mathscr{E}(x)}{dx^2}=0$
with boundary conditions $\mathscr{E}(0)=T_l$ and $\mathscr{E}(1)=T_r$. This equation has a linear solution for the temperature profile in the thermodynamic limit:
$\mathscr{E}(x)=T_l+(T_r-T_l)x$.
For large $L$, using $e_i=\mathscr{E}(i/L)$ in eq. (\ref{current_i_2}), we find
$J_i^{(e)} \simeq \frac{2}{L} (d \mathscr{E}(x) / dx )_{x=i/L} \simeq \frac{2}{L}~(T_l-T_r)$, which signifies
the validity of Fourier's law. The fact that Fourier's law is valid in our model is not surprising since the dynamics is diffusive and the model falls in the class of
models known as ``gradient type'' \cite{kipnis-book, Spohn90}. These are models for which one has a local continuity equation at the discrete level and the current
can be written as a discrete gradient of some function of the local energy. Indeed, in our model local current can be written as a discrete
gradient of the local energy.

Although the local average energy profile is linear in the $L \to \infty$ limit, there are deviations from this linearity for finite size systems.
We can compute these deviations by solving the recursion relation (\ref{e-prof-recur}) exactly for arbitrary $L$. 
Eq. (\ref{e-prof-recur}) can be written in a matrix form 
\begin{eqnarray}
&&~~~~~\mathbb{T}{\bf e}={\bf b},~~~ 
\text{where},~~{\bf e}=(e_1,e_2,...,e_L),~~{\bf b}=(-3T_l,-T_l,0,...,0,-T_r,-3T_r), ~~~\text{and}~~\mathbb{T}_{r,s}=u_{r-s}
\end{eqnarray}
with $u_{0}=-6,~u_{1}=u_{-1}=2,~u_{2}=u_{-2}=1$ and $u_{\nu}=0,~|\nu| \geq 3 $. Using the theory of inverse of Toeplitz matrices from \cite{Trench85}
one can find $\mathbb{T}^{-1}$, which provides the energy profile ${\bf e}= \mathbb{T}^{-1}{\bf b}$. To find $\mathbb{T}^{-1}$, one uses
the method of characteristic polynomials where, one starts with a trial solution $e_i \sim w^i$. Inserting this trial solution in eq. (\ref{e-prof-recur}) 
we get the polynomial equation
$w^4+2w^3-6w^2+2w+1=0$ whose roots are $w_1=1,~w_2=1$, $w_3=-(2-\sqrt{3})$ and $w_4=-(2+\sqrt{3})$. As a result the general solution for $e_i$
can be written as $e_i= a + bi + c w_3^i + d w_4^i$, where $a,b,c$ and $d$ are unknown constants determined by the boundary conditions given in
eq. (\ref{e-prof-recur}). After some algebraic manipulations one finds that
the energy profile is given by
\begin{eqnarray}
 e_{i}&=&T + \frac{\Delta T}{2} B(L-1,i),~~i=1,2,...,L,~~~\text{with}~~ T=\frac{T_l+T_r}{2},~~\Delta T= T_l-T_r~~\text{and}\label{e-profile} \\
 B(n,i)&=&\frac{(3 n-6 i) \left [26 a_-(n)-15 \sqrt{3} a_+(n)\right ]-9~[a_-(n-i)-a_-(i)]+5 \sqrt{3}~[a_+(n-i)-a_+(i)]}
{n~\left[78a_-(n)-45 \sqrt{3} a_+(n)\right]+189 a_-(n)-109 \sqrt{3}
   a_+(n)+2 \sqrt{3}} \label{B_n-i} \\
\text{where},&& a_+(x)=(-1)^x[(2-\sqrt{3})^x+(2+\sqrt{3})^x],~~ a_-(x)=(-1)^x[(2-\sqrt{3})^x-(2+\sqrt{3})^x].
\end{eqnarray}
This is our first main result. For large $n$, one can easily see that
\begin{eqnarray}
&&B(n,r)\simeq \left(1-\frac{2r}{n}\right)~\times~\left[1+\frac{3279+1889\sqrt{3}}{n}\right]^{-1} +(-1)^n\frac{(2+\sqrt{3})^{-n}}{n}~\times~
\left(\text{Boundary~terms}\right) \label{large-n-approx}
\end{eqnarray}
which implies the linear temperature profile given in eq. (\ref{profiles}) for large $L$. Here we define the local temperature by the average local energy :
$T_i=e_i$. In fig. \ref{fig3} we compare the solution $e_{i}=T + \frac{\Delta T}{2} B(L-1,i),~~i=1,2,...,L$ with the results of 
numerical simulations where good agreement is evident. We observe that for large system size the energy profile is 
indeed in the scaling form $e_i=\mathscr{E}(i/L)$ and is dominantly linear. 
The local current $J_i^{(e)}$ inside the bulk can be obtained by inserting $e_i$ from eq. (\ref{e-profile}) in eq. (\ref{current_i_2}).

%

In this paper we mainly consider the baths at the two ends of the system as having different temperatures
but zero average momenta. Let us however make some remarks on the setup where they have different average momenta as well. 
This can be achieved by considering a shifted Gaussian as the bath momentum distribution, instead of the distributions in eqs. (\ref{ht-bth-l}) and (\ref{ht-bth-r}).
In this case there will also be momentum current in addition to the energy current, and this current will also support an average momentum profile
across the system.
Following exactly same steps for obtaining the energy profile in eq. (\ref{e-prof-recur}), one can also obtain a similar recursion relation
for the average momentum $m_i=\langle p_i \rangle$ :
\begin{equation}
 6m_i=m_{i-2}+2m_{i-1}+2m_{i+1}+m_{i+2},~~i=1,2,...,L-1,L;~~\text{with}~~m_{-1}=m_0=M_l,~~m_{L+1}=m_{L+2}=M_r, \label{mom-prof-recur}
\end{equation}
where $M_l$ and $M_r$ are the average momenta of the left and the right reservoirs respectively. 
Here we observe that the recursion relation satisfied by $m_i$ is the same as the one satisfied by $e_i$ in  eq. (\ref{e-prof-recur}), 
and is independent of the energy drive.
One can immediately write the solution as  $m_{i}= (M_l+M_r)/2 + (M_l-M_r)/2~ B(L-1,i),~~i=1,2,...,L$ where the function
$B(n,r)$ is given in eq. (\ref{B_n-i}). In the large $L$ limit, the average momentum profile is also well approximated by a linear profile
$m_i=\mathscr{M}(i/L)$ with $\mathscr{M}(x)=M_l+(M_r-M_l)x$, which provides the average momentum current as $J_i^{(p)} \simeq \frac{2}{L}~(M_l-M_r)$.

\section{Finite size corrections }
\label{F-S-correc}
\noindent
 In the preceding section we have computed an exact expression of the local temperature profile for arbitrary system size $L$. In the large $L$ limit, we
have seen that the local temperature profile can be well approximated by a linear profile $e_i=\mathscr{E}(i/L)$ where $\mathscr{E}(x)=T_l+(T_r-T_l)x$.
This allows us to write the local equilibrium distribution as
$g_i(p_i)= \frac{e^{-p_i^2/2e_i}}{\sqrt{2\pi e_i}}$ and the joint distribution of the momenta as
$P^{(L)}(\{p_i\}) \simeq \prod g_i(p_i) [~1 + O(1/L) ~]$ at the leading order. However, for finite size systems there are deviations in the local marginal
distribution as well as in the joint distribution. In this section we look at such deviations as finite size corrections.
It is clear that to find such deviations from the local equilibrium distribution, we need to compute various correlations.
Since the local distribution is Gaussian at the leading order, the natural
quantity to measure the deviation from the Gaussianity is the kurtosis $K_i=\langle p_i^4 \rangle-\langle p_i^2 \rangle^2$.
Additionally, deviation from the factorized joint distribution
can be quantified by various correlations. In this section, we compute the kurtosis $K_i$,
the two site momentum-momentum correlations $\mathcal{P}_{ij}= \langle p_ip_j \rangle-\langle p_i \rangle \langle p_j \rangle$ and the energy-energy correlations
${C}_{ij}= \langle p_i^2p_j^2 \rangle-\langle p_i^2 \rangle \langle p_j^2 \rangle$ as finite size corrections to the LE state.

To evaluate $K_i$, $\mathcal{P}_{ij}$ or $C_{ij}$, we need to perform averages over the NESS in which the joint distribution satisfies
the master equation (\ref{SS-mstr-eq}). Using this equation we obtain recursion relations involving correlations among neighboring sites. As a result
we will have a set of linear equations involving $K_i,~\mathcal{P}_{ij},~C_{ij}$ as well as all other fourth-order (in momentum) correlation functions
$\bar{C}_{i,j}=\langle p_i^3 p_j \rangle_c$, $C_{i,j,k}=\langle p_i^2 p_j p_k \rangle_c$ and $C_{i,j,k,l}=\langle p_i p_j p_k p_l \rangle_c$. Here the subscript $c$
represents the cumulant. Note that these linear equations do not involve higher order correlation functions in momentum, and hence they form a closed set of equations.
For a given system of size $L$, there are $\mathcal{O}(L^4)$ such linear equations.
Solving them analytically for arbitrary $L$
is a rather difficult task. Solving them numerically is also hard for large $L$ (say of order $L=50$ ).
However from numerical simulations we find that the correlations $C_{i,j}$ decay as $\sim 1/L$ for large $L$ and they have good scaling under the 
transformation $x=i/L,~y=j/L$. In this section we compute these scaling forms of the correlations using an appropriate continuum description.

We begin our analysis by first considering a simpler multi-channel variant of the model. In this variant, when the number of channels is very large, the correlations between 
sites vanish, \emph{i.e.} $C_{ij}=0,~\mathcal{P}_{ij}=0$. For this variant we calculate the kurtosis and show that it scales as
$K_i \sim 1/L^2$. Next, we return to our original model, where correlations are present,
and show that they enhance the deviation from Gaussianity and $K_i \sim 1/L$, in this case.


\subsection{The multi-channel model}
\label{MF}
\begin{figure}[t]
\includegraphics[scale=0.4]{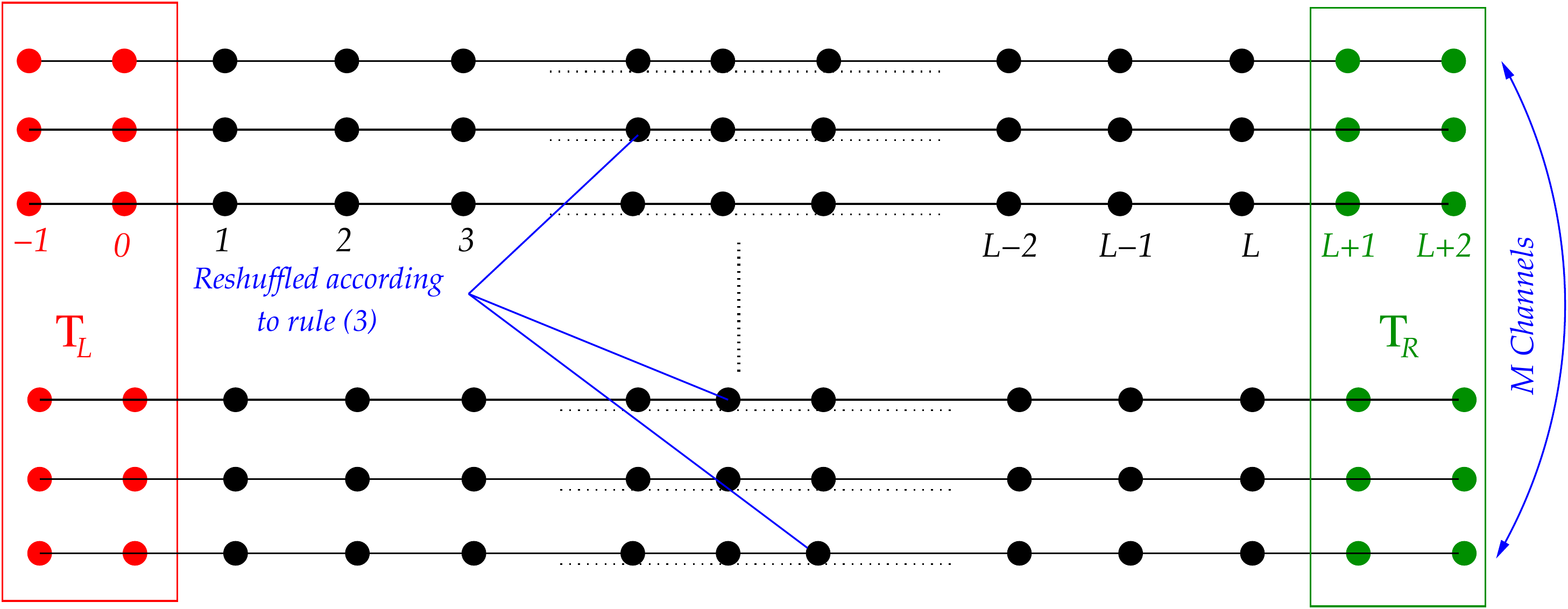}
\caption{ (Color Online) Schematic diagram of the multi channel model.}
\label{MF-model}
\end{figure}

\begin{figure}[t]
\includegraphics[scale=0.4]{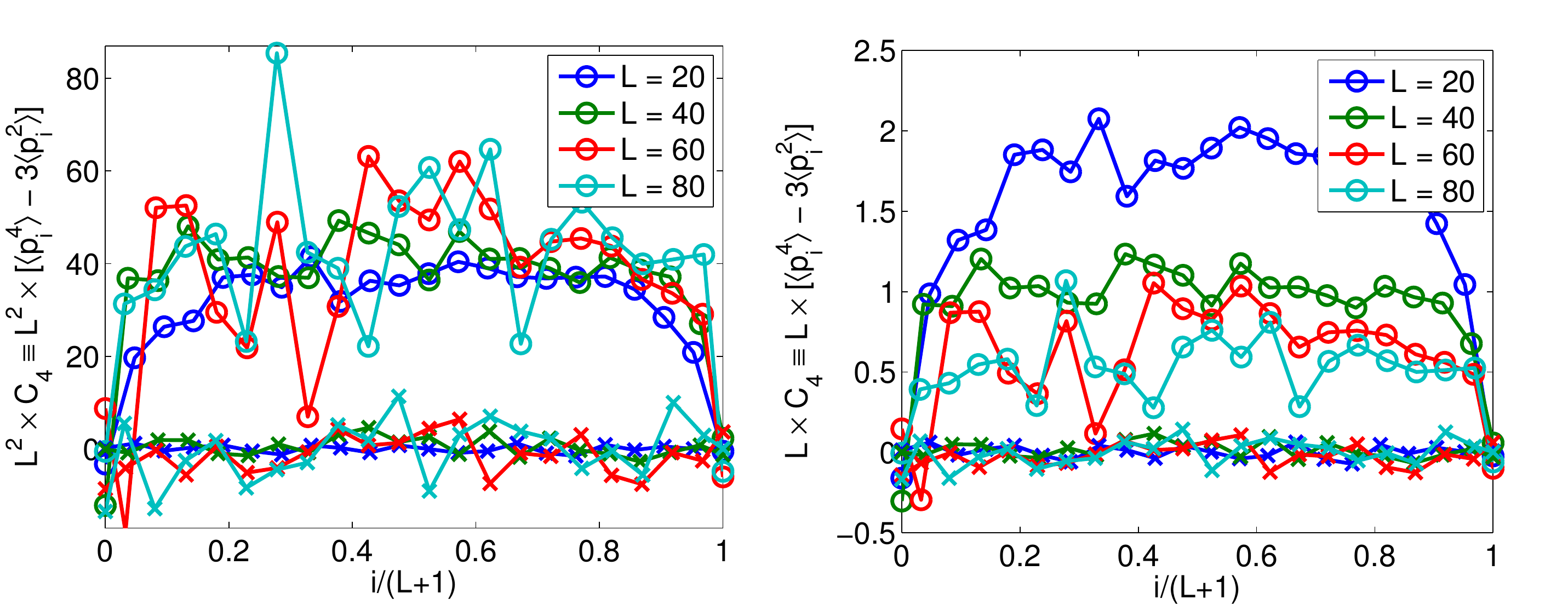}
\caption{ (Color Online) Scaling of kurtosis obtained from simulation in the multi-channel model for $T_l=4$ and $T_r=2$ (circles) and for equilibrium $T_r=T_l=2$ (crosses).
In the left panel we plot $L^2~K_i$ vs. $i/(L+1)$
whereas, for comparison, we plot $L~K_i$ vs. $i/(L+1)$ in the right panel. From these plots it is clear that the kurtosis $K_i$ for the multi-channel
case scales as $\sim 1/L^2$.  In equilibrium $K_i=0$.}
\label{MF-kurt}
\end{figure}
\noindent
Here we consider a modified model in which different sites are uncorrelated in some appropriate limit. This model can also be viewed as a mean
field version of the original model defined above. To define the modified model let us consider $M$ channels
(see fig. \ref{MF-model}) , \emph{i.e.} $M$ identical copies of the one-dimensional lattice described in sec. \ref{Model}. As before, each site carries
a momentum $p_{i,a}$, where the subscript $i=-1,...,L+2$ denotes the location of the site along lattice in $a$th channel and
the superscript $a=1,...,M$ denotes the channel number. At each simulation step, a site $0 \leq i \leq L+1$ is selected
at random. In addition, three channels $1 \leq a,b,c \leq M$ are selected independently at random. Then the three momenta $p_{i-1,a}$, $p_{i,b}$
and $p_{i+1,c}$ are mixed according to the collision rule (\ref{rules}), as before. In the $M \to \infty$ limit, correlations between consecutive
sites are expected to become negligible. In the simulations we consider $M=100$ and we run most of our simulations for $10^6$ time steps
(unless otherwise specified). Here one time step consists of $(L+2)M$ collisions.

We have studied the average local energy, current and the kurtosis as a function of site index for different system sizes. We observe that the
local energy profile is again given by eq. (\ref{e-profile}) as in the correlated case. This is a result of the overall energy conservation.
The average local current scale as $\sim 1/L$ and thus again satisfies the Fourier's Law. In fig. \ref{MF-kurt}, we plot
the kurtosis $K_i$ as function of $i/(L+1)$. In the left panel we plot $L^2~K_i$ vs. $i/(L+1)$ whereas for comparison we also plot $L~K_i$ vs. $i/(L+1)$ in
the right panel. From these plots we can evidently see that the kurtosis in the mean field case scales as $~1/L^2$. We have also compared the data presented
in fig. \ref{MF-kurt} with other data sets corresponding to different combinations of $T_l$ and $T_r$ (not provided here) and have found that $K_i$ does not
depend on the mean temperature $(T_l+T_r)/2$ but is proportional to $(T_l-T_r)^2$.

Let us now make an effort to understand the above observations from an analytical point of view.
We start by approximating the joint distribution of the three momenta $P^{(3)}(p_{i-1,a},p_{i,b},p_{i+1,c})$ appearing in eq. (\ref{NESSpdf})
as the product of the three marginal distributions, \emph{i.e.}
\begin{eqnarray}
 P^{(3)}(p_{i-1,a},p_{i,b},p_{i+1,c}) = P^{(1)}(p_{i-1,c})~ P^{(1)}(p_{i,b}) ~P^{(1)}(p_{i+1,c}). \label{mean-field-approx}
\end{eqnarray}
From now onwards we omit explicit appearances of the channel labels $a,~b,~c$ for convenience.
Naturally the above approximation makes it easier to compute the kurtosis $K_i$. Note that the equation for the local energy $e_i$ remains the
same as in eq. (\ref{e-prof-recur}). {This is because the average local energy $e_i$ depends only on the energies of the neighboring sites but not
on the correlations between different sites. This fact can be easily seen from the computation of $\mathcal{I}_i$ in eq. (\ref{mcalI}). }

We now present two alternative methods to obtain an equation for the kurtosis $K_i$: by a direct derivation from the master equation, and by a perturbative 
calculation of the marginal distribution $P^{(1)}(p)$. The first method allows, in principle, an exact determination of the finite size kurtosis. While the 
second holds only asymptotically when $\Delta T/L$ is small, it can be used to find also higher order momentum cumulants. 

Method 1: To obtain a difference equation satisfied by $K_i$'s, we compute $\langle p_i^4 \rangle = \int_{-\infty}^{\infty} dp_i ~p_i^4~P^{(1)}(p_i)$
using eq. (\ref{NESSpdf}) and the approximation (\ref{mean-field-approx}).
Once again integrating the $p$ variables first with the help of eq. (\ref{transformation}), we get steady state average of polynomials of order $q^4$
involving $q_{i\pm 2},~q_{i \pm1}$ and $q_{i}$. Now subtracting $3\langle p_i^2 \rangle^2$ from $\langle p_i^4 \rangle;~~\forall i$, we get
\begin{eqnarray}
&&~~~~~~~~~~~~186 K_i-19(K_{i-2}+2K_{i-1}+2K_{i+1}+K_{i+2}) = \Lambda_0(i),~~~\text{for}~i=1,...L,~~\text{where}, \label{Kurtosis-mean-field} \\
&&\Lambda_0(i) = 24 \left[ (e_{i-2}e_{i-1}+e_{i-1}e_i+e_ie_{i-2}) + (e_{i-1}e_{i}+e_{i+1}e_i+e_{i-1}e_{i+1})
+(e_{i}e_{i+1}+e_{i}e_{i+2}+e_{i+1}e_{i+2})\right] \label{lambda-0} \\
&& ~~~~~~~~~~~~~~~~~~~~~~~~~~~~+57~(e_{i-2}^2+2e_{i-1}^2+2e_{i+1}^2+e_{i+2}^2) - 558~e_i^2. \nonumber
\end{eqnarray}
$\Lambda_0(i)$, given in eq. (\ref{lambda-0}), is a function of the local average energies of the sites.
The above equation should be solved with the following boundary conditions :
$ K_{-1}=K_0=0$, and $K_{L+1}=K_{L+2}=0$.
For any given $L$, one can in principle solve this recursion relation numerically quite easily but it is quite cumbersome to solve it analytically.
However, we are interested in the large $L$ behavior. Hence we use the following large $L$ approximation
$e_i \simeq T+\frac{\Delta T}{2} \left(1-2 \frac{i}{L}\right)$ [see eq. (\ref{large-n-approx})]
of the energy profile to compute the $\Lambda_0(i)$'s in eq. (\ref{Kurtosis-mean-field})
and find that 
\begin{equation}
 \Lambda_0(i) \simeq 756~ \Delta T^2 / L^2,
\end{equation}
for all $i$. This immediately implies that $K_i$ scales as $\sim \Delta T^2/L^2$.

Method 2: This method involves perturbatively correcting the local equilibrium ansatz for 
the single point distribution $P^{(1)}(p_i)$. At the zeroeth order in $\Delta T/L$, the local equilibrium distribution is given by
the following Gaussian distribution
\begin{eqnarray}
 p^{(1)}(p_i) &=& g(p_i,T_i),~~
\text{where},~~g(p,T) = \frac{\text{exp}\left(-\frac{p^2}{2T}\right)}{\sqrt{2 \pi T}},\label{ansatz}
\end{eqnarray}
and $T_i$ is given in eq. (\ref{profiles}). Noting the fact that $T_i=T+\alpha$ (where $\alpha$ is small and is of order $(\Delta T/L)$ when $\Delta T/L$ is small)
we observe $g(p,T+\alpha)$ has the following expansion in powers of $\alpha$ as
\begin{eqnarray}
 g(p,T+\alpha) \simeq g(p,T)\left[1+\frac{\alpha}{\sqrt{2}T}\phi_2\left(\frac{p}{\sqrt{T}}\right)
+ \frac{\alpha^2\sqrt{6}}{4T^2} \phi_4\left(\frac{p}{\sqrt{T}}\right) + ....\right]. \label{g-expnsn}
\end{eqnarray}
It turns out that the functions $\phi_n(x)$s are the eigenfunctions of the ``collision'' operator $\mathcal{K}$, \emph{i.e.} they satisfy \cite{Ma83}
\begin{eqnarray}
&&\int dp' dp'' dqdq'dq''g(p',T)g(p'',T) ~\mathcal{K} ~\phi_n\left(\frac{q}{\sqrt{T}}\right) = \frac{2\pi}{3\sqrt{3}}(1-\gamma_n)
\phi_n\left(\frac{p}{\sqrt{T}}\right), \nonumber \\
&&~~~~~~~~~~~~~~~~~~~
~~\text{where},~~\gamma_n= \left[1-3\left( \frac{d^n}{d x^n}e^{x/3}I_0(-2x/3)\right)_{x=0}\right].\nonumber
\label{mcalK-op}
\end{eqnarray}
and $I_0(y)$ is zeroeth order modified Bessel function.
In \cite{Ma83} it was shown that $\phi_n\left(x\right)=\frac{1}{\sqrt{2^nn!}}~H_n(x/\sqrt{2})$ where $H_n(y)$ is the $n$th
order Hermite polynomials.
The first few eigenvalues and eigenfunctions are
\begin{eqnarray}
\label{ex-egf-egv}
\begin{array}{cc}
\phi_0\left(x\right)=1, & \gamma_0=0 \\
\phi_1\left(x\right)=x, & \gamma_1=0 \\
\phi_2\left(x\right)=\frac{1}{\sqrt{2}} \left(x^2 -1\right), & \gamma_2=0 \\
\phi_3\left(x\right)=\frac{1}{\sqrt{6}} x\left(x^2 -3\right), & \gamma_3= \frac{2}{9} \\
\phi_4\left(x\right)=\frac{1}{2\sqrt{6}} \left(x^4 -6x^2 +3\right), &
\gamma_4= \frac{8}{27} \\
\phi_6\left(x\right)=\frac{1}{12\sqrt{5}} \left(x^6 -15x^4 +45x^2 -15\right), &
\gamma_6= \frac{34}{81} \\
\end{array}
\end{eqnarray}
The fact that first three eigenvalues are zero is a consequence of the conservation of the particle number, momentum and energy.

Our aim is to improve over the local equilibrium distribution and that is done by assuming $p^{(1)}(p_i)= g(p_i,T_i)~\Psi(p_i,T_i)$ such that
the function $\Psi(p_i,T_i) \to 1$ as $L\to \infty$ or $\Delta T \to 0$ or both. Suggested by the expansion in eq. (\ref{g-expnsn}),
we argue that function $\Psi(p_i,T_i)$ also has an expansion in the basis of the eigenfunctions $\phi_n(x)$ of the operator $\mathcal{K}$.
This means
\begin{eqnarray}
 \Psi(p_i,T_i) &=& 1 + A_i~\phi_2(p_i/\sqrt{T_i}) + B_i~\phi_4(p_i/\sqrt{T_i})+D_i~\phi_6(p_i/\sqrt{T_i})+.... ,\label{ansatz1}
\end{eqnarray}
where the coefficients $A_i$, $B_i,...$ have to be determined from $\langle p_i^2 \rangle$, $\langle p_i^4 \rangle$ etc. Note that no odd order
eigenfunctions appear in the expansion since the joint distribution $p^{(L)}(\{p_i\})$ is symmetric under global inversion of all momenta
$\{p_i\} \to \{-p_i\}$. Using
$p^{(1)}(p_i)= g(p_i,T_i)~\Psi(p_i,T_i)$, one computes $\langle p_i^2 \rangle=T_i+A_i$ implying $A_i=0$. On the other hand,
computing $\langle p_i^4 \rangle$, one finds $K_i=\langle p_i^4 \rangle-3\langle p_i^2 \rangle^2=2\sqrt{6}T_i^2B_i$, 
which with help of eq. (\ref{Kurtosis-mean-field}) implies
 $B_i \simeq \Delta T^2/L^2$. As a result we have $p^{(1)}(p_i)= g(p_i,T_i)~[1+B_i~\phi_4(p_i/\sqrt{T_i}) + o(1/L^2)]$.
To check whether this form of $p^{(1)}(p_i)$ actually solves the eq. (\ref{NESSpdf}) to order $\mathcal{O}(1/L^2)$ we first insert this form in eq. (\ref{NESSpdf}). Next,
using the approximation (\ref{mean-field-approx}) we get after some manipulations and simplification
\begin{eqnarray}
&&B_i \phi_{4}\left(\frac{p}{\sqrt{T}}\right) = \frac{1}{3}~ \int dp'dp''~dqdq'dq''~\mathcal{K}~
g\left(p',T_i\right)~g\left(p'',T_{i}\right)~\times \label{eq-A-i} \\
&&~~~~~~~~~~~~~~~~~~\left[ \left(\frac{3\alpha^2\sqrt{6}}{T_i^2} + G_i \right) \phi_{4}\left(\frac{q}{\sqrt{T_i}}\right)
+\frac{3\alpha^2}{2T_i^2} \phi_{2}\left(\frac{q'}{\sqrt{T_i}}\right)\phi_{2}\left(\frac{q''}{\sqrt{T_i}}\right)\right]. \label{chk-sol} \\
&&\text{where,}~~~~G_i~=~B_{i-2}~+~2~B_{i-1}~+~3~B_i~+~2~B_{i+1}~+~B_{i+2} \nonumber
\end{eqnarray}
\begin{figure}[t]
\includegraphics[scale=0.35]{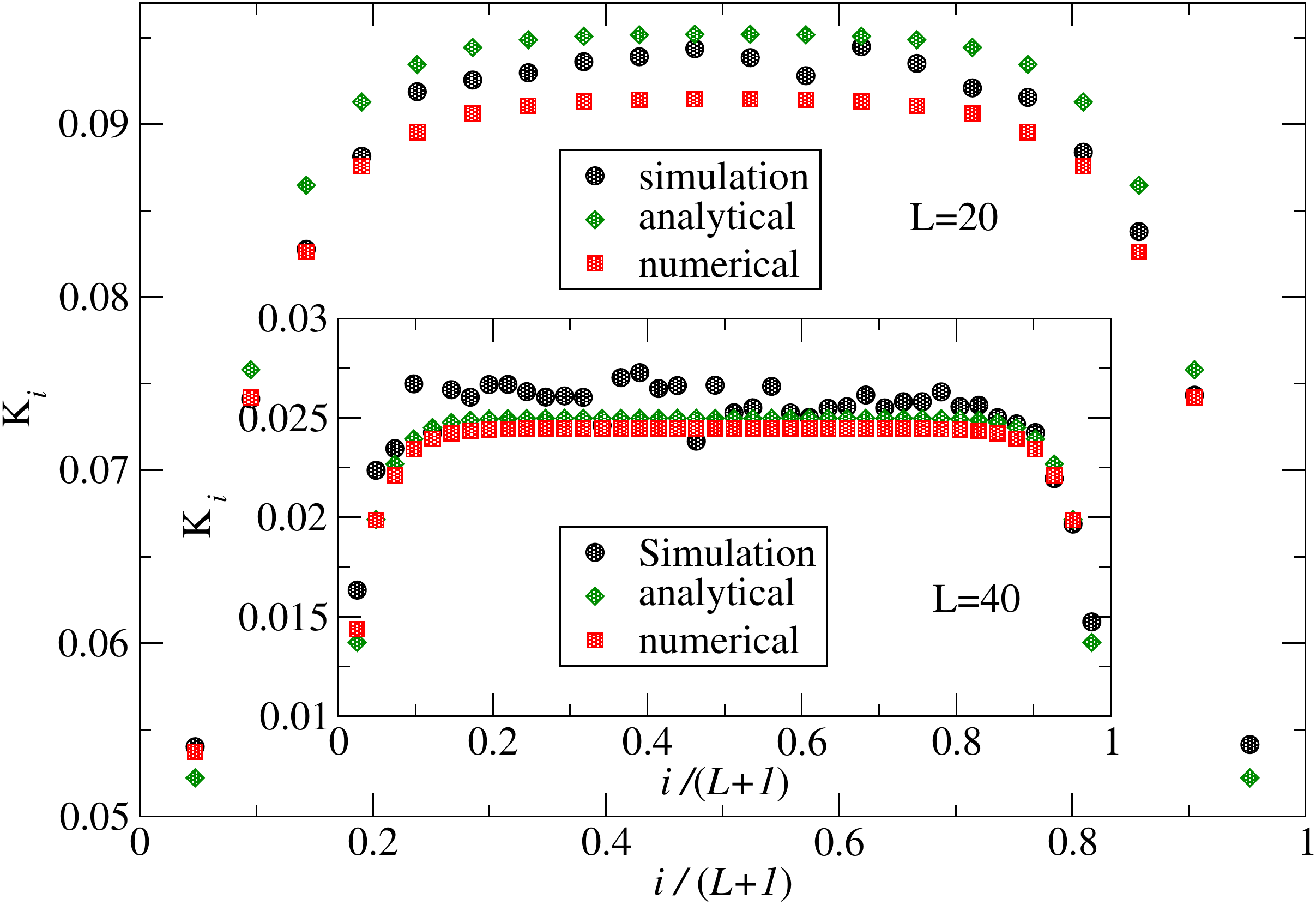}
\caption{ (Color Online) Comparison of kurtosis profile of the multi channel model obtained from simulation, numerical solution and the 
approximate analytical expression (\ref{K-prof-LA-2})
 for $L=20$ and $40$.
The deviations from the analytical result
are due to the linear approximation of the
energy profile where we have neglected the next order term \emph{i.e.} $\mathcal{O}(1/L)$, and due to not enough statistics.
The temperatures for this plot are $T_l=4.0$ and $T_r=2.0$.
}
\label{MF-kurt-comparison}
\end{figure}
To proceed further we use the following relation ,
\begin{eqnarray}
&& \int dp'dp''~dqdq'dq''~\mathcal{K}~g\left(p',T_i\right)~g\left(p'',T_{i}\right)~\phi_{2}\left(\frac{q'}{\sqrt{T_i}}\right)
~\phi_{2}\left(\frac{q''}{\sqrt{T_i}}\right) =
\frac{2\pi}{\sqrt{3}}~\frac{4\sqrt{6}}{81}~\phi_4\left(\frac{p}{\sqrt{T_i}}\right) \nonumber
\end{eqnarray}
which can be proved using eq. (\ref{mcalK-op}).
Using this relation in eq. (\ref{chk-sol}), we immediately see that $p^{(1)}(p_i)= g(p_i,T_i)~[1+B_i~\phi_4(p_i/\sqrt{T_i}) ]$ satisfies eq. (\ref{NESSpdf}) if
$B_i$ satisfies
\begin{equation}
186~B_i-19~(~B_{i-2}~+~2~B_{i-1}~+~2~B_{i+1}~+ B_{i+2}~) \simeq \frac{378}{\sqrt{6}T^2}~\frac{\Delta T^2}{L^2~},
\end{equation}
which is the same as eq. (\ref{Kurtosis-mean-field}) in large $L$ limit because $K_i\simeq 2\sqrt{6}~T^2~B_i$.

The solution of equation (\ref{Kurtosis-mean-field}) is given by
\begin{equation}
K_i \simeq \frac{21}{2}~ \frac{\Delta T^2}{L^2}
~\left[ 1-\frac{\sinh \left(\sqrt{\frac{12}{19}}~(L-i)\right)
+\sinh \left(\sqrt{\frac{12}{19}} ~i\right)}{\text{sinh}\left(\sqrt{\frac{12}{19}} ~L\right) }\right], \label{K-prof-LA-2}
\end{equation}
to leading order in $(1/L)$. The second term in the square brackets above is displayed to show how fast $K_i$ decays from its bulk value to zero as
$i$ approaches to the boundary.
We observe that the kurtosis $K_i$ scales as $\sim \Delta T^2/L^2$ and does not depend on the mean temperature $T=(T_l+T_r)/2$. In the bulk, $K_i$ is
more or less flat \emph{i.e.} independent of $i$, with a sharp drop to zero over a range of order one from the boundary.
In fig. \ref{MF-kurt-comparison} we compare the kurtosis profile given in eq. (\ref{K-prof-LA-2})
with the same profile obtained from direct numerical simulation for $L=20$ and $L=40$.
We see that the agreement is good although there are some visible differences. They appear mainly because of
lack of statistics and the linear approximation of the energy profile where we have neglected
the next order term \emph{i.e.} terms of $\mathcal{O}(1/L)$.

In summary, we find that when the different sites of our model are uncorrelated, and the finite size correction to the local equilibrium
Gaussian distribution, quantified by the kurtosis scales as $\sim \Delta T^2/L^2$.
Moreover, we find an explicit expression of the
profile of the kurtosis $K_i$ as a function of the site index $i$. This is our second main result.
One can perform similar analysis to find the coefficients of the higher order functions in eq (\ref{ansatz1}).
In the next section we consider the original problem where the
different sites are correlated and there we will see that the corresponding finite size corrections {\it do not} scale as $\sim \Delta T^2/L^2$,
but rather as $\sim \Delta T^2/L$.

\subsection{The correlated case}
\label{corr-only-e-drive}
\noindent
\begin{figure}[t]
\includegraphics[scale=0.35]{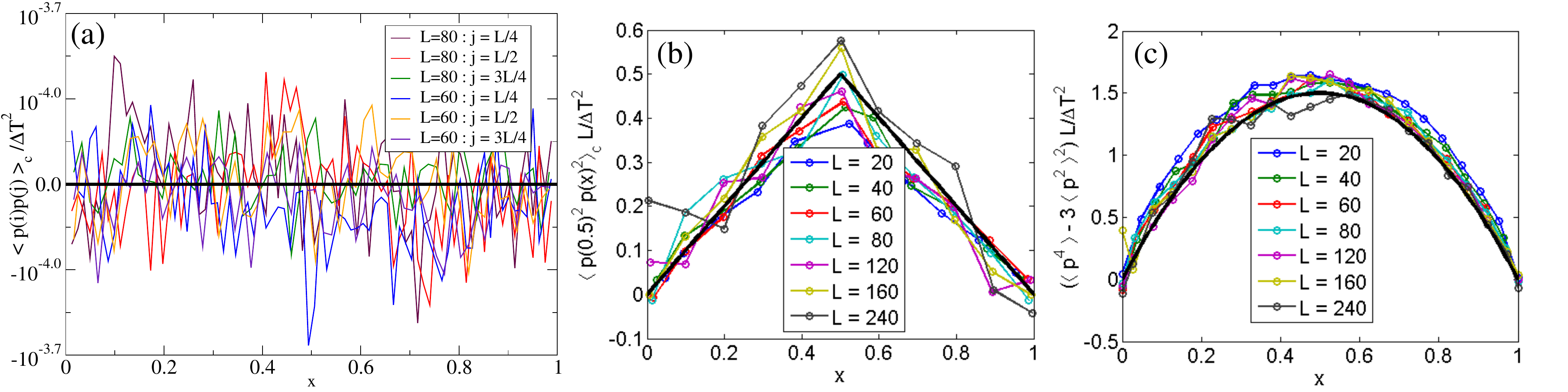}
\caption{ (Color Online) Scaling of $C_{i,j}$ and $K_i$ obtained from simulation for $T_l=2.0$ and $T_r=1.0$.
In panel (a) we plot $\mathcal{P}_{i,j}~L/\Delta T^2$ vs. $x=i/(L+1)$ and find $\mathcal{P}_{i,j} \simeq 0$.
In panel (b) we plot
$C(i,L/2)~L/\Delta T^2$ vs. $x=i/(L+1)$ and in panel (c) we plot $K_i~L^2/\Delta T^2$ vs. $x=i/(L+1)$.
From the data collapse in plots (b) and (c) we see that, $C(i,L/2)$ and $K_i$ scale as $\sim 1/L$.
}
\label{kurt-corr-scaling-ori}
\end{figure}
In this section, we consider the original problem of a single channel model where different sites are correlated. Let us start by presenting
our simulation results. We performed simulations to compute the momentum-momentum correlations
$\mathcal{P}_{i,j}= \langle p_i p_j \rangle-\langle p_i \rangle\langle p_j \rangle$,
the energy-energy correlations
$C_{i,j}=\langle p_i^2 p_j^2 \rangle- \langle p_i^2\rangle\langle p_j^2\rangle$ and also the kurtosis
$K_i=\langle p_i^4\rangle- 3~\langle p_i^2\rangle^2$ for different system sizes. The simulations were run over $10^8$ time steps where each time step
consists of $L+2$ collisions. In fig. \ref{kurt-corr-scaling-ori}(a) we plot $\mathcal{P}_{i,j}$ vs. $x=i/(L+1)$ and find that it is zero.
In fig. \ref{kurt-corr-scaling-ori}(b) and \ref{kurt-corr-scaling-ori}(c), we plot the
rescaled correlation functions $(L/\Delta T^2)~C(L/2,i)$ and rescaled kurtosis $(L/\Delta T^2)~K_{i}$ against $x=i/(L+1)$ respectively, and find that data for
different system sizes collapse on a single function in both cases. This data collapse suggests that both $C_{i,j}$ and $K_i$ scale as
\begin{equation}
 K_i \simeq \frac{1}{L}~\mathscr{K}\left ( \frac{i}{L}\right),~~C_{i,j} \simeq \frac{1}{L}~\mathscr{C}\left ( \frac{i}{L}, \frac{j}{L}\right),
~~~\text{for~large~}~L,
\label{ecp-sclng-form}
\end{equation}
in contrast to the $\sim 1/L^2$ scaling in the mean field case.
In the following we provide an analytical approach to understand these findings.

We start by computing $\mathcal{P}_{i,j}= \langle p_i p_j \rangle-\langle p_i \rangle\langle p_j \rangle$. 
Since there is no momentum drive, the joint distribution of momenta $P^{(L)}(\{p_i\})$ is
symmetric under global momentum reversal : $p_i \to -p_i,~~\forall i$. Therefore one can readily see that
$\langle p_i \rangle=0,~\forall i$ and hence $\mathcal{P}_{i,j}= \langle p_i p_j \rangle$ in this case.
To compute $\langle p_i p_j \rangle$, we multiply both sides of (\ref{SS-mstr-eq}) by $p_ip_j$ and integrate over all the momenta. Finally, we get
\begin{eqnarray}
&&12\mathcal{P}_{i,j} = (\mathcal{P}_{i-2,j}+2\mathcal{P}_{i-1,j}+2\mathcal{P}_{i+1,j}+\mathcal{P}_{i,j})
+ (\mathcal{P}_{i,j-2}+2\mathcal{P}_{i,j-1}+2\mathcal{P}_{i,j+1}+\mathcal{P}_{i,j+2}),~~~1 \leq i < j \leq L,~~|i-j|\geq 3, \nonumber \\
&&10\mathcal{P}_{i,i+2}=\mathcal{P}_{i-2,i+2} + 2 \mathcal{P}_{i-1,i+2} + 2 \mathcal{P}_{i+1,i+2} + 2 \mathcal{P}_{i,i+1} + 2  \mathcal{P}_{i,i+3}
+ \mathcal{P}_{i,i+4},~~~1 \leq i < L, \label{mom-mom-corr}\\
&&8\mathcal{P}_{i,i+1}=\mathcal{P}_{i-2,i+1} + \mathcal{P}_{i-1,i} + 2 \mathcal{P}_{i-1,i+1} + 2 \mathcal{P}_{i,i+2} + \mathcal{P}_{i,i+3}
+ \mathcal{P}_{i+1,i+2},~~~1 \leq i < L-1, \nonumber
\end{eqnarray}
where $\mathcal{P}_{\mu,\nu}=0$ if any one of the indices $\mu$ or $\nu$ falls on the boundary sites : $-1,0,L+1,L+2$. The equation for $i=j$ is given in
eq. (\ref{e-prof-recur}) as $e_i=\langle p_i^2 \rangle$. Since, the two point correlation  $\mathcal{P}_{i,j}$ does not depend on the one-point function
$e_i=\langle p_i^2 \rangle$, the set of eqs. (\ref{mom-mom-corr}) is homogeneous.
As a result, the solution of eqs. (\ref{mom-mom-corr}) is $\mathcal{P}_{i,j}=0,~~\forall~i \neq j$.

Next we consider various correlations of order $p^4$. Note once again that all odd order correlations are zero because the joint distribution
$P^{(L)}(\{p_i\})$ is symmetric under global momentum reversal.
There are five types of
correlations at order $p^4$ : one 1-point correlation $K_i=\langle p_i^4\rangle- 3~\langle p_i^2\rangle^2$, two 2-point correlations
$C_{i,j}=\langle p_i^2 p_j^2 \rangle_c$ and $\bar{C}_{i,j}=\langle p_i^3 p_j \rangle_c$, one 3-point correlation
$C_{i,j,k}=\langle p_i^2 p_j p_k \rangle_c$ and one 4-point correlation $C_{i,j,k,l}=\langle p_i p_j p_k p_l \rangle_c$.
Here the subscript $c$ once again represents cumulant, \emph{e.g.} $\langle p_i^2 p_j^2 \rangle_c= \langle p_i^2 p_j^2 \rangle-\langle p_i^2 \rangle \langle p_j^2 \rangle$.
Starting from eq. (\ref{SS-mstr-eq}), one can write equations for the five types of correlations and
find that they satisfy $S_L=\frac{L}{24}(L^3+18 L^2 -25 L +6)$ coupled linear equations.
To see how they are coupled let us write for example the equation for the $K_i$. To do this we use
eq. (\ref{NESSpdf}) and follow the same procedure as done in computing $K_i$ in the mean field case.
The main difference is that eq. (\ref{mean-field-approx}) no longer holds. As was done in the evaluation of $\mathcal{I}_i$ in eq. (\ref{mcalI}),
we use the transformation (\ref{transformation})
and integrate the $p$ variables first to obtain average of polynomials of order $q^4$ involving different combinations of $q$-momenta from different sites,
which provides the following difference equation satisfied by $K_i$ for $1\leq i \leq L$ :
\begin{eqnarray}
 &&186~ K_i=19~[~ K_{i-2} + 2 K_{i-1} +2K_{i+1} + K_{i+2}~]~+~I_{\mathcal{A}}(i)~+~ \Lambda_0(i) \nonumber \\
&&~~~~~~~~~~~~~~~~~~~~+24~[~C_{i-2,i-1} +C_{i-2,i} + 2C_{i-1,i} + C_{i-1,i+1}+2C_{i,i+1}+ C_{i+1,i+2}+C_{i,i+2}~ ] \label{4th-mom-corr} \\
&&\text{where} \nonumber \\
&&\Lambda_0(i) = 24 \left[ (e_{i-2}e_{i-1}+e_{i-1}e_i+e_ie_{i-2}) + (e_{i-1}e_{i}+e_{i+1}e_i+e_{i-1}e_{i+1})
+(e_{i}e_{i+1}+e_{i}e_{i+2}+e_{i+1}e_{i+2})\right] \label{lambda-0-1} \\
&& ~~~~~~~~~~~~~~~~~~~~~~~~~~~~+57~(e_{i-2}^2+2e_{i-1}^2+2e_{i+1}^2+e_{i+2}^2) - 558~e_i^2. \nonumber
\end{eqnarray}
Here any correlation with boundary sites : $(-1, 0, L+1, L+2)$ is taken to be zero.
The term $I_{\mathcal{A}}(i)$ depends on $\bar{C}_{i,j}$ and $C_{i,j,k}$, and is given explicitly in eq. (\ref{mcalIA}) in appendix A.
Note that in the absence of any
correlations, \emph{i.e.} when correlations like $C_{i,j}$,
$\bar{C}_{i,j}$, $C_{i,j,k}$ and $C_{i,j,k,l}$ are zero,
the above equation reduces to the mean field equation (\ref{Kurtosis-mean-field}). However, these correlations are
not identically zero in general.
Following a similar procedure used for obtaining equation for $K_i$, one can also write equations for other correlations. For example,
multiplying both sides of eq. (\ref{SS-mstr-eq}) by $p_i^2p_j^2$ and integrating over the momenta, one can get the equations satisfied by
$C_{i,j}=\langle p_i^2 p_j^2 \rangle_c$ for different $i$ and $j$:
\begin{eqnarray}
&&240C_{i,i+1} = [27C_{i-2,i+1}+15C_{i-1,i}+42C_{i-1,i+1}+42C_{i,i+2}+15C_{i+1,i+2}+27C_{i,i+3}] \label{corr-eq-apndx-2} \\
&&~~~~~~~~~~~~~~~~~~~~~~~+~4[K_{i-1}+2K_{i}+2K_{i+1}+K_{i+2}]~+~I_{\mathcal{B}_1}(i) ~+~\Lambda_1(i),~~~~~~~~~~~~~~~1 \leq i < L, \label{C-aprox-1} \\
&&282C_{i,i+2} = [27C_{i-2,i+2}+54C_{i-1,i+2}+42C_{i+1,i+2}+42C_{i,i+1}+ 54C_{i,i+3}+27C_{i,i+4}] \label{corr-eq-apndx-3} \\
&&~~~~~~~~~~~~~~~~~~~~~~~~~~~~~~~~~~~~~~~+ 4[K_{i}+K_{i+1}+K_{i+2}]~+~I_{\mathcal{B}_2}(i) +\Lambda_2(i),~~~~~~~~~~1 \leq i < L-1, \nonumber\\
&&12C_{i,j} = [C_{i-2,j}+2C_{i-1,j}+2C_{i+1,j}+C_{i+2,j}] + [C_{i,j-2}+2C_{i,j-1}+2C_{i,j+1}+C_{i,j+2}],\label{corr-eq-apndx-1} \\
&&~~~~~~~~~~~~~~~~~~~~~~~~~~~~~~~~~~~~~~~~~~~~~~~~~~~~~~~~~~~~~~~~~~~~~~~~~~~~~~~~~~~~~~~~~1 \leq i < j \leq L,~|i-j|\geq 3,~ \label{C-aprox-2}\\
&&\text{where,} \nonumber \\
&&~~~~~~~~\Lambda_1(i) = 27e_{i-2}e_{i+1}+15e_{i-1}e_{i}+42e_{i-1}e_{i+1}+42e_{i}e_{i+2}+15e_{i+1}e_{i+2}+27e_{i}e_{i+3}] \label{lambda-1} \\
&&~~~~~~~~~~~~~~~~~~~~~~~~~~~~~~~~~~~+~12[e_{i-1}^2+2e_{i}^2+2e_{i+1}^2+e_{i+2}^2]- 240 e_{i}e_{i+1} \nonumber \\
&&\text{and}\nonumber \\
&&~~~~~~~~~~~~~~\Lambda_2(i) = [27e_{i-2}e_{i+2}+54e_{i-1}e_{i+2}+42e_{i+1}e_{i+2}+42e_{i}e_{i+1} \label{lambda-2} \\
&&~~~~~~~~~~~~~~~~~~~~~~~~~+ 54e_{i}e_{i+3}+27e_{i}e_{i+4}] + 12[e_{i}^2+e_{i+1}^2+e_{i+2}^2]-282e_{i}e_{i+2}\nonumber
\end{eqnarray}
As before any correlations with boundary sites
are taken to be zero. The terms $I_{\mathcal{B}_{1,2}}$ are given in eqs. (\ref{mcalIB1}) and (\ref{mcalIB2}), whereas
the inhomogeneous terms $\Lambda_{1,2}(i)$ are given in eqs. (\ref{lambda-1}) and (\ref{lambda-2}). One can similarly write equations satisfied by
$\bar{C}_{i,j}$, $C_{i,j,k}$ and $C_{i,j,k,l}$. As these equations are rather lengthy and numerous, we present a few such equations in appendix
\ref{appendix-2} [see eqs. (\ref{p3p}), (\ref{pp2p}) and (\ref{pppp})]. In fact there are $\mathcal{O}(L^4)$ rather lengthy linear coupled equations 
for a given $L$. One can in principle find all the possible correlations
in the system for a given size $L$ and temperatures $T_l$ and $T_r$ by solving this system of linear equations, but solving them
analytically for arbitrary $L$ is highly
non-trivial. Even solving them numerically for moderately large $L$ is difficult. However as we are looking for solutions of $K_i$ and $C_{i,j}$
in the scaling form given in eq. (\ref{ecp-sclng-form}), we find that these correlations can be obtained
 by considering a proper continuum description in the $L \to \infty$ limit.

Since we are interested in the large $L$ limit we
simplify the inhomogeneous terms $ \Lambda_0(i)$, $\Lambda_1(i)$ and $\Lambda_2(i)$ using the large $L$ linear
approximation of the energy profile : $e_i \simeq T_l - \Delta T~(i/L)$. We find $ \Lambda_0(i) \simeq 756~\frac{\Delta T^2}{L^2}$,
$ \Lambda_1(i) \simeq 18~\frac{\Delta T^2}{L^2}$ and $ \Lambda_2(i) \simeq -72~\frac{\Delta T^2}{L^2}$.
\begin{figure}[t]
\includegraphics[scale=0.35]{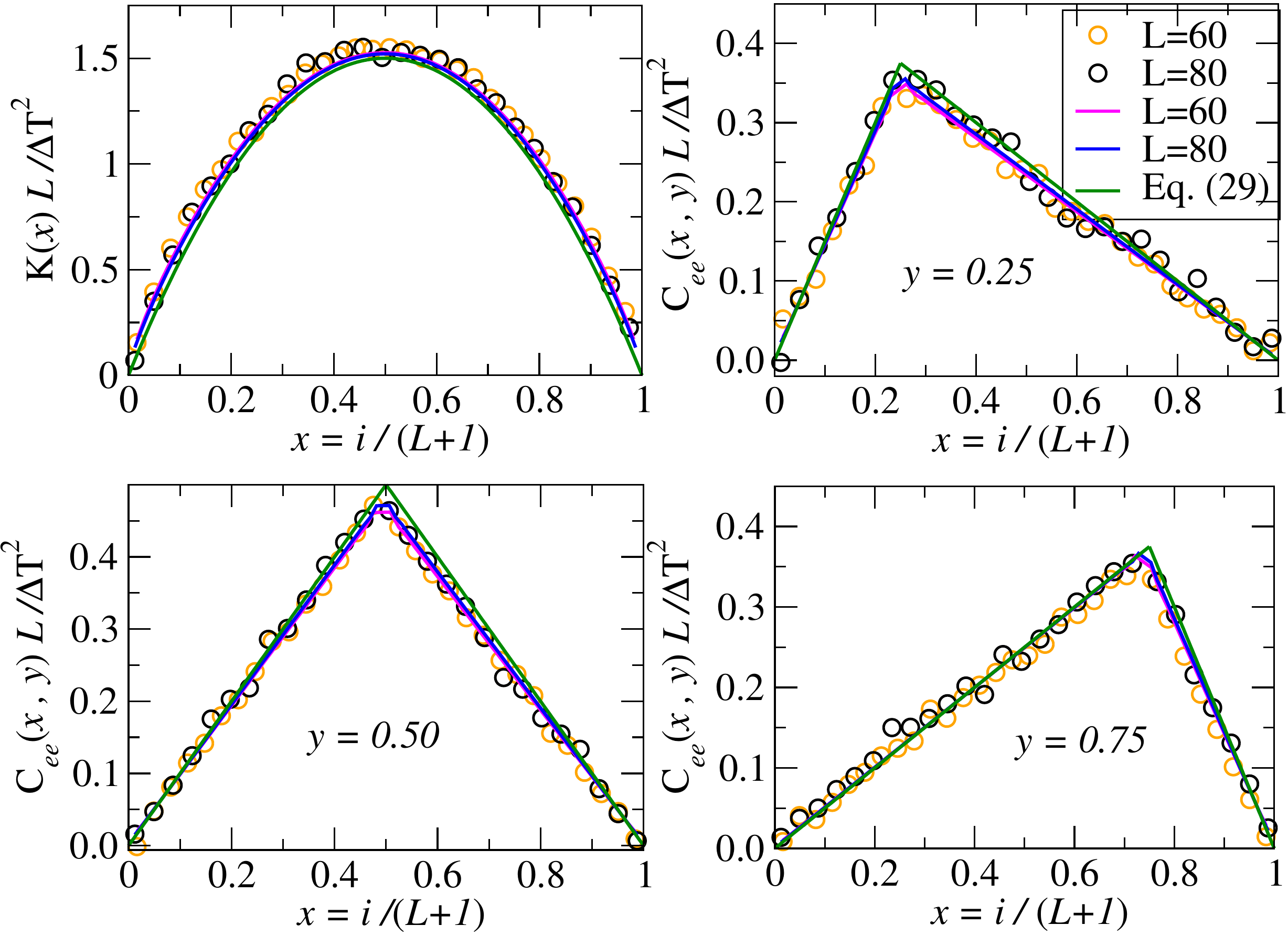}
\caption{ (Color Online) The correlation $C_{i,j}$ and the kurtosis $K_i$ obtained from numerical solution of eqs. (\ref{4th-mom-corr})-(\ref{corr-eq-apndx-1})
neglecting contributions from $\bar{C}_{i,j}$ and $C_{i,j,k}$. We compare this numerical result with that obtained from
Monte-Carlo simulation. In all four figures, open circles ($\circ$) represent data obtained from simulations whereas the red
({\color{red}--}) and the blue ({\color{blue}--}) lines represent results of the numerical calculations.
The green line ({\color{green}--}) represents the theoretical result in eq. (\ref{sol-c-ee}) with .
}
\label{corr-compre-simu-num-guess}
\end{figure}
As a result the source terms of the difference equations (\ref{4th-mom-corr}), (\ref{corr-eq-apndx-2}) and (\ref{corr-eq-apndx-3}),
provided by the inhomogeneous terms,
are of order $\sim \Delta T^2/L^2$ and are independent of $i$. In addition, there are two other types of contributions at
$\mathcal{O}\left(\frac{\Delta T^2}{L^2}\right)$ in the source terms. The first one comes from correlations like
$\bar{C}_{i,j}$ and $C_{i,j,k}$ through the terms $I_{\mathcal{A}}$, $I_{\mathcal{B}_{1}}$ and $I_{\mathcal{B}_{2}}$ and the second
contribution results from the fact that the kurtosis itself
can have such a term in the next order \emph{i.e.}
$K_i = \frac{1}{L}~\mathscr{K}\left ( \frac{i}{L}\right) + \alpha \left(\frac{i}{L}\right) \frac{\Delta T^2}{L^2}$ where
$\alpha\left(x\right)$ could be a smooth slowly varying function. We later show that
the contributions from the later two possibilities
cancel out exactly in the continuum limit, although they appear explicitly in individual equations at the discrete level. 
Hence the net source or the inhomogeneity in eqs. (\ref{4th-mom-corr}), (\ref{corr-eq-apndx-2}) and (\ref{corr-eq-apndx-3}) are of order $\frac{\Delta T^2}{L^2}$.

If we now expect our solutions to be of the scaling form (\ref{ecp-sclng-form})
then the only possible way to equate with a source of order $\sim \Delta T^2/L^2$ is
through the first derivatives of the scaling functions, which will provide an extra $1/L$.
This means, after inserting the scaling forms (\ref{ecp-sclng-form}) in eqs. (\ref{4th-mom-corr})-(\ref{corr-eq-apndx-3}), that 
if we Taylor expand them around $x=y=i/L$ as powers of $1/L$, we should not have any non zero term at order $\mathcal{O}(1/L)$.
This implies $\mathscr{K}(x)=3 \mathscr{C}(x,y)|_{|x-y|\to 0} + \mathcal{J}$, where $\mathcal{J}$ possibly contains the order $\sim 1/L$ 
contributions of $\bar{C}_{i,j}$ and $C_{i,j,k}$ that appear in the equations 
of $K_i$ and $C_{i,j}$ through $I_{\mathcal{A}}$, $I_{\mathcal{B}_{\pm 1}}$ and $I_{\mathcal{B}_{\pm 2}}$. It turns out that these contributions too 
cancel out (to order $\sim 1/L$) as we show in Appendix \ref{appendix-3}. 
Hence at order $\mathcal{O}(1/L)$, $\mathcal{J}=0$ and we have the following relation between the two scaling functions $\mathscr{K}(x)$ and $\mathscr{C}(x,y)$,
\begin{equation}
 \mathscr{K}(x)=3 \mathscr{C}(x,y)|_{|x-y|\to 0},~~\text{for}~(x,y) \in [0,1]\times [0,1]. \label{K-eq-3C}
\end{equation}

The above equation yields a simple relation between the kurtosis $K_i$ and the two-site energy-energy correlation $C_{i,j}$ but it does not provide $C_{i,j}$ or
$\mathscr{C}(x,y)$ itself.
We now use eq. (\ref{K-eq-3C}) to derive a differential equation for $\mathscr{C}(x,y)$ in the continuum limit.
To obtain this equation, it is convenient to rewrite the set of equations involving $K_i$ and $C_{i,j}$, valid in both the upper half triangle
and the lower half triangle, in terms of a single function
$C_{i,j}$ in the following way : Taking care of possible double counting of the terms coming from the equations on the $i=j$ line, we join
the set of equations from the upper and lower half triangles and then substitute $K_i = 3 C_{i,i},~\forall~i$. We get,
\begin{eqnarray}
&&~~~~~~~~~[\mathcal{L}C](i,j) = \mathcal{A}(C)\delta_{i,j} + \mathcal{B}_1(C)\delta_{i+1,j}
+  \mathcal{B}_{-1}(C)\delta_{i-1,j}
+ \mathcal{B}_2(C)\delta_{i+2,j} +  \mathcal{B}_{-2}(C)\delta_{i-2,j}, \label{eq-ful-plane} \\
&&\text{where}, \nonumber \\
&& ~~~~~~~~~[\mathcal{L}C](i,j) = 12 C_{i,j} - (~C_{i-2,j}+2C_{i-1,j}+2C_{i+1,j}+C_{i+2,j}~)- (~C_{i,j-2}+2C_{i,j-1}+2C_{i,j+1}+C_{i,j+2}~), \nonumber
\end{eqnarray}
$~~~~~$ and
\begin{eqnarray}
&&\mathcal{A}(C) = 57[C_{i-2,i-2}+2C_{i-1,i-1}+2C_{i+1,i+1}+C_{i+2,i+2}] -546C_{i,i} +\Lambda_0(i) + I_{\mathcal{A}}(i)
,\nonumber \\
&&~~~~~~~~~+12C_{i-2,i-1}+11C_{i-2,i}+22C_{i-1,i}+12C_{i-1,i+1}+22C_{i,i+1}+11C_{i,i+2}+12C_{i+1,i+2} \nonumber \\
&&~~~~~~~~~+12C_{i-1,i-2}+11C_{i,i-2}+22C_{i,i-1}+12C_{i+1,i-1}+22C_{i+1,i}+11C_{i+2,i}+12C_{i+2,i+1}, \nonumber \\
&& \nonumber \\
&&\mathcal{B}_1(C) = 26C_{i-2,i+1}+40C_{i-1,i+1}+40C_{i,i+2}+26C_{i,i+3}+14C_{i-1,i}+14C_{i+1,i+2} \nonumber \\
&&~~~~~~~~~~~-228C_{i,i+1}+12C_{i-1,i-1}+22C_{i,i}+22C_{i+1,i+1}+12C_{i+2,i+2} +\Lambda_1(i)+I_{\mathcal{B}_{1}}(i), \nonumber \\
&&\nonumber \\
&&\mathcal{B}_{-1}(C) = 26C_{i,i-3}+40C_{i,i-2}+40C_{i+1,i-1}+26C_{i+2,i-1}+14C_{i-1,i-2}+14C_{i+1,i} \nonumber \\
&&~~~~~~~~~~~-228C_{i,i-1}+12C_{i-2,i-2}+22C_{i-1,i-1}+22C_{i,i}+12C_{i+1,i+1} +\Lambda_1(i)+I_{\mathcal{B}_{-1}}(i), \nonumber \\
&&\nonumber \\
&&\mathcal{B}_{2}(C)=26C_{i-2,i+2}+52C_{i-1,i+2}+40C_{i+1,i+2}+40C_{i,i+1}+52C_{i,i+3}+26C_{i,i+4} \nonumber \\
&&~~~~~~~~~~~-270C_{i,i+2} + 11C_{i,i}+12C_{i+1,i+1}+11C_{i+2,i+2} +\Lambda_2(i)+I_{\mathcal{B}_{2}}(i), \nonumber \\
&&\nonumber \\
&&\mathcal{B}_{-2}(C)=26C_{i,i-4}+52C_{i,i-3}+40C_{i,i-1}+40C_{i-1,i-2}+52C_{i+1,i-2}+26C_{i+2,i-2} \nonumber \\
&&~~~~~~~~~~~-270C_{i,i-2} + 11C_{i,i}+12C_{i-1,i-1}+11C_{i-2,i-2} +\Lambda_2(i)+I_{\mathcal{B}_{-2}}(i). \nonumber
\end{eqnarray}
The functions $ \Lambda_0(i)$, $\Lambda_1(i)$ and $\Lambda_2(i)$ are given explicitly
in eqs. (\ref{Kurtosis-mean-field}), (\ref{corr-eq-apndx-2}) and (\ref{corr-eq-apndx-3}) whereas explicit expressions of the functions $I_{\mathcal{A}}(i)$, 
$I_{\mathcal{B}_{\pm 1}}(i)$ and $I_{\mathcal{B}_{\pm 2}}(i)$ are given in Appendix \ref{appendix-2}.
Now in the above equation (\ref{eq-ful-plane}) we
use the scaling form $C_{i,j} = \frac{1}{L}\mathscr{C}\left( \frac{i}{L},\frac{j}{L}\right)$ and make the following variable transformation
$x=i/L$ and $y=j/L$. Under this transformation and in $L \to \infty$ limit one finds
$\sim-(6/L^3)~(\partial_x^2 + \partial_y^2)~\mathscr{C}(x,y) + \mathcal{O}(1/L^4)$ on the left hand side (lhs) and
$=\frac{1}{L}~\delta(y-x)~[\mathcal{A}+\mathcal{B}_1+\mathcal{B}_{-1}+\mathcal{B}_{2}+\mathcal{B}_{-2}]_{L\to \infty}$ on the right hand side (rhs).
Using the linear approximation for $e_i$ one finds that the contribution from the $\Lambda$-terms to the source on the rhs is given by
$[\Lambda_0(i)+2\Lambda_1(i)+2\Lambda_2(i)]_{L\to \infty} \simeq 648~\frac{\Delta T^2}{L^2}$. On the other hand, the contribution from correlations such as  
$\bar{C}_{i,j}$ and $C_{i,j,k}$ to the source on the rhs comes only through the combination 
$[I_{\mathcal{A}}(i)+I_{\mathcal{B}_1}(i)+I_{\mathcal{B}_{-1}}(i)+I_{\mathcal{B}_2}(i)+I_{\mathcal{B}_{-2}}(i)]$. From their explicit expressions in eqs. 
(\ref{mcalIA}), (\ref{mcalIB1}), (\ref{mcalIBm1}), (\ref{mcalIB2}) and (\ref{mcalIBm2}) one can show, after some cumbersome manipulations, that 
\begin{equation}
 [I_{\mathcal{A}}(i)+I_{\mathcal{B}_1}(i)+I_{\mathcal{B}_{-1}}(i)+I_{\mathcal{B}_2}(i)+I_{\mathcal{B}_{-2}}(i)]=0. \label{mcalIeq0}
\end{equation}
Therefore in the continuum limit, the 
correlations $\bar{C}_{i,j}$ and $C_{i,j,k}$ have no contribution to the source on the rhs. In a similar way one can show that the contribution from terms such as  
$\alpha(i/L)~\Delta T^2/L$ at the second leading order in the kurtosis also cancel out exactly.

The rest of the contribution to the source on rhs comes from the discontinuities of first derivatives of $\mathscr{C}(x,y)$ across $x=y$ line.
To compute this contribution,
let us denote $[\partial_x~\mathscr{C}(x,y)]_{x\to y^+}$ by $\delta_x^b$ and $[\partial_x~\mathscr{C}(x,y)]_{x\to y^-}$ by $\delta_x^t$ and similarly,
$[\partial_y~\mathscr{C}(x,y)]_{y\to x^+}$ by $\delta_y^t$ and $[\partial_y~\mathscr{C}(x,y)]_{y\to x^-}$ by $\delta_y^b$. Using this notation,
one can show that
\begin{eqnarray}
\mathcal{A}(C) &\simeq& \frac{1}{L^2}~[~68~(\delta_x^b-\delta_x^t) ~+~ 68~(\delta_y^t-\delta_y^b)~] + 756~\frac{\Delta T^2}{L^2} , \nonumber \\
\mathcal{B}_1(C)&\simeq& \frac{1}{L^2}~[~-92~\delta_x^t ~+~24~\delta_y^t ~+~\frac{34}{3}\partial_x \mathscr{K}(x)~]~+~ 18~\frac{\Delta T^2}{L^2}, \nonumber \\
\mathcal{B}_{-1}(C)&\simeq& \frac{1}{L^2}~[~92~\delta_x^b ~-~24~\delta_y^b ~-~\frac{34}{3}\partial_x \mathscr{K}(x)~~]~+~ 18~\frac{\Delta T^2}{L^2}, \label{ABD-dis} \\
\mathcal{B}_2(C)&\simeq& \frac{1}{L^2}~[~-64~\delta_x^t ~-~4~\delta_y^t ~+~\frac{34}{3}\partial_x \mathscr{K}(x)~]~-~ 72~\frac{\Delta T^2}{L^2}, \nonumber \\
\mathcal{B}_{-2}(C)&\simeq& \frac{1}{L^2}~[~64~\delta_x^b ~+~4~\delta_y^b ~-~\frac{34}{3}\partial_x \mathscr{K}(x)~]~-~ 72~\frac{\Delta T^2}{L^2}.\nonumber
\end{eqnarray}
Using the above expressions we find that the rhs $=\frac{1}{L^3}~\delta(y-x)~[~648~\Delta T^2~+~224(\delta_x^b-\delta_x^t)+88(\delta_y^t-\delta_y^b)~]$
in the leading order. The leading terms on lhs and rhs are of the same order and equating them we get the following differential equation:
$-6~\nabla^2~\mathscr{C}(x,y)= \delta(y-x)~[~648~\Delta T^2~+~224(\delta_x^b-\delta_x^t)+88(\delta_y^t-\delta_y^b)~]$, where
$\nabla^2=\partial_x^2 + \partial_y^2$. Since by definition the correlation $C_{i,j}$ is symmetric under $i \leftrightarrow j$ interchange,
the scaling function $\mathscr{C}(x,y)$ is also symmetric under $x \leftrightarrow y$ interchange. This means, $\mathscr{C}(x,y)$ should also satisfy
the differential equation $-6~\nabla^2~\mathscr{C}(x,y)= \delta(y-x)~[~648~\Delta T^2~+~224(\delta_y^b-\delta_y^t)+88(\delta_x^t-\delta_x^b)~]$, which is
obtained by interchanging $x$ and $y$ in the previous differential equation. Hence after proper symmetrization we have
\begin{equation}
-6~(\partial_x^2 + \partial_y^2)~\mathscr{C}(x,y)= \delta(y-x)~[~648~\Delta T^2~+~156(\delta_y^t-\delta_x^t)~-156(\delta_y^b-\delta_x^b)~],
\end{equation}
which, under the following variable transformation $u=y-x,~v=y+x$, can be rewritten as
\begin{equation}
 -12~(\partial_u^2+\partial_v^2)~\mathscr{C}(u,v)= \delta(u)~[~648~\Delta T^2~+~312(\partial_u \mathscr{C}(u,v)|_{u\to 0^+}~
-~\partial_u \mathscr{C}(u,v)|_{u\to 0^-})~].
\end{equation}
Integrating both sides over $u$ from $u=0^-$ to $u=0^+$, one finds
$(\partial_u \mathscr{C}(u,v)|_{u\to 0^+}-\partial_u \mathscr{C}(u,v)|_{u\to 0^-})= -2 \Delta T^2$. As a result the above equation becomes
$(\partial_u^2+\partial_v^2)~\mathscr{C}(u,v) = -2 \Delta T^2 \delta(u)~$ which under the reverse transformation $x=(v-u)/2,~y=(v+u)/2$
becomes
\begin{equation}
 (\partial_x^2 + \partial_y^2)~\mathscr{C}(x,y) = - 4~\Delta T^2~\delta(x-y). \label{corr-conti}
\end{equation}
The boundary conditions for the above equation are given by
$\mathscr{C}(0,y)=\mathscr{C}(1,y)=\mathscr{C}(x,0)=\mathscr{C}(x,1)=0$, because at the boundary the system is connected
to the heat reservoirs which individually maintain independent Gaussian distributions of momenta. Equation (\ref{corr-conti})
can be interpreted as an electrostatic problem of finding the potential created by the charges distributed uniformly along the diagonal of
a square of length one with zero potential at the boundaries. Similar analogies with electrostatic problems have been used in \cite{TS14} while 
computing long range density-density correlation in a diffusive lattice gas with simple exclusion interaction.
Recently, a general theory called the macroscopic fluctuation theory for diffusive systems has been developed \cite{Bertini02, Bertini14,ADO},
using which one also can obtain the same differential equation in the large system size limit.
The solution of eq. (\ref{corr-conti}) is given by
$\mathscr{C}(x,y) = 2~\Delta T^2~x(1-y)$ for $0 \leq x \leq y \leq 1$. As a result we have,
\begin{eqnarray}
\label{sol-c-ee}
&&C_{i,j} \simeq \frac{1}{L}~\mathscr{C}\left ( \frac{i}{L}, \frac{j}{L}\right)~~\text{where},~~~
\mathscr{C}(x,y) = 2~\Delta T^2~\begin{cases}
                                                 & ~x(1-y),~~~x>y \\
                                                 & ~y(1-x),~~~y>x
                                                \end{cases},  \\
&&\text{and}~~~~~~~K_i \simeq \frac{1}{L}~\mathscr{K}\left ( \frac{i}{L}\right)~~~\text{where},~~~\mathscr{K}(x) =  6~\Delta T^2~x(1-x).\label{sol-K}
\end{eqnarray}
We see that the kurtosis $K_i$ and the correlation $C_{i,j}$ are both of order $\mathcal{O}(\Delta T^2/ L)$ for large $L$, and at that order even 
macroscopically distant sites are correlated. 
This long range correlation at order $\sim 1/L$ is a result of broken detailed balance in the NESS. 
Such long range correlations in non-equilibrium systems have been studied in other contexts 
\cite{Dorf94, Zarate06, Nicolis84, Spohn83, Garrido60}. 
Starting from a microscopic description, similar expressions of long-range correlations have been computed in
chemical systems \cite{Nicolis84}, using a lattice gas automaton approach \cite{suarez96} and in the random halves models \cite{Lin07, Larralde09}.
This $\mathcal{O}(1/L)$ scaling of correlation stems from the fact that in the large
space and time limit (diffusive scaling limit), both the mean and the variance of the fluctuating local current scale as $\sim 1/L$ 
\cite{Derrida07, Derrida10, Bertini07, Lee95} 
and this fluctuation of current controls all correlations in the hydrodynamic limit. This behavior has been shown in various
microscopic models, such as the simple symmetric exclusion model \cite{Derrida07, Derrida10}, the KMP model \cite{Bertini07}.

Along with the numerical simulation we have also attempted to solve equations (\ref{C-aprox-1})-(\ref{C-aprox-2}) 
numerically using Mathematica and compared the results obtained via 
the three methods : simulations, numerical calculation and the theoretical result in eqs. (\ref{sol-c-ee}) and (\ref{sol-K}). 
Details of the simulations have already been presented
in the beginning of this subsection. For the numerical solution method we solve
a smaller set ($\mathcal{O}(L^2)$~) of equations obtained by neglecting contributions from $\bar{C}_{i,j}$ and $C_{i,j,k}$ in
eqs. (\ref{4th-mom-corr})-(\ref{corr-eq-apndx-1}). One can in principle solve all the equations involving $\bar{C}_{i,j}$, $C_{i,j,k}$ and  $C_{i,j,k,l}$,
but, as mentioned earlier, there are $\mathcal{O}(L^4)$ such equations and solving them is rather demanding computationally.
In fig. \ref{corr-compre-simu-num-guess} we compare this numerical result along with the same obtained from direct numerical simulations and with the
theoretical results (\ref{sol-c-ee}) and (\ref{sol-K}) and see a very nice agreement.

\section{Conclusion}
\label{conclu}
In this paper we have studied energy transport through a one dimensional stochastic model system conserving both energy and momentum in the bulk.
To define a local dynamics that conserves both total energy and momentum, we considered a three particle collision interaction similar to that considered
in \cite{Basile06, Bernardin08, Basile07, Basile09}. This system, when connected to two reservoirs at its two ends, reaches a stationary state 
where there is a flow of energy current from
high temperature to low temperature. We have found that this current satisfies Fourier's law and the associated local temperature profile 
is linear in the thermodynamic limit.
For finite size systems there are deviations from this linear behavior which we have computed exactly.
We have also found that, in the large system size limit, the stationary state of the system can locally 
be described by the equilibrium Gaussian distribution associated to the
local temperature. 
In the second part of the paper, we have computed the kurtosis and spatial correlations of the momenta from different sites for large systems. 
We have shown that these correlations are long ranged.

First, we have considered a simpler variant of our original model, the multi channel model, 
where all the correlations are zero, \emph{i.e} $\mathcal{P}_{ij}=0$, $C_{ij}=0$ etc. For this case, the deviation from the local equilibrium Gaussian distribution,
quantified by kurtosis, decays as $K_i \sim \frac{1}{L^2}$ for large but finite $L$. On the other hand for the original fully correlated model
we have found that, interestingly, both the two point energy-energy correlation $C_{ij}$ and the kurtosis $K_i$ decay as $\sim \frac{1}{L}$.
Moreover, in the later case, $C_{ij}$ and $K_i$ scales as $C_{ij} = \frac{1}{L} \mathscr{C}(i/L,~j/L)$ and $K_{i} = \frac{1}{L} \mathscr{K}(i/L)$ for large $L$, 
respectively. By deriving an appropriate continuum limit from the
discrete equations satisfied by various correlations, we have shown that $\mathscr{K}(x)=3\mathscr{C}(x,~x)$ and $\mathscr{C}(x,~y)$ satisfies a differential
equation, which can be interpreted as an electrostatic problem of finding the potentials in a unit square with uniform charge distribution along the diagonal and with zero
potential at the boundaries. Solving this differential equation has yielded explicit expressions of $\mathscr{C}(x,~y)$ and hence $\mathscr{K}(x)$.

In this paper, we have mostly considered the case where the two reservoirs have different temperatures but zero average momentum. As a result, there is only energy drive
but no momentum drive. It will be interesting to study this model when both drives are present. In this situation, correlations with odd powers in momentum $p$
will also be non zero. Computing these correlations will provide more insight into the nature of the non-equilibrium steady state.
In fact, finding the joint distribution of all the momenta would be ideal but this is a highly difficult task.
A relatively easier quantity to compute would probably be the single-site marginal distribution.

Various studies have shown that the fluctuations and correlations in non-equilibrium steady state created in different geometries behave differently. 
For example, in stochastic models with a single conserved quantity (like exclusion processes in one dimension), one finds different correlations 
in the following two situations: when the NESS is
created by boundary drive (like boundary temperature difference) in an open geometry, and when the NESS is created by a local drive in a ring geometry \cite{Bordineau}. 
In our three particle collision model, it will also be interesting to look at such differences.

Finally, computing the large deviation function of the local energy and momentum profile in the thermodynamic limit is also an interesting open problem.

\section{Acknowledgement}
We thank Julien Cividini and Sanjib Sabhapandit for helpful discussions. The support of the Israel Science Foundation (ISF) and of the Minerva
Foundation with funding from the Federal German Ministry for Education and Research is gratefully acknowledged.
\appendix
\section{Explicit expressions of the equations satisfied by various correlations}
\label{appendix-2}
\noindent
In this appendix we present some of the equations satisfied by $K_i=\langle p^4_i \rangle_c$,
 $C_{ij}=\langle p^2_ip_j^2 \rangle_c$,
$\bar{C}_{ij}=\langle p_i^3 p_j \rangle_c$, \\
 $C_{ijk}=\langle p_i^2 p_j p_k \rangle_c$ and $C_{ijkl}=\langle p_i p_j p_k p_l \rangle_c$.
Here the subscript $c$ represents cumulant. These equation are obtained from the NESS master equation (\ref{SS-mstr-eq}). For example to obtain equation for
$\langle p_i^3 p_j \rangle$, one multiplies both sides of eq. (\ref{SS-mstr-eq}) by $p_i^3 p_j$ and then integrates over all momenta $\{p_i\}$. Following this
procedure we get equations for $K_i,~C_{ij},~\bar{C}_{ij},~C_{ijk}$ and $C_{ijkl}$. Below we present some such equations :

\noindent\rule{18cm}{0.6pt}
\begin{eqnarray}
 &&186~ K_i=19~[~ K_{i-2} + 2 K_{i-1} +2K_{i+1} + K_{i+2}~]~+~I_{\mathcal{A}}(i)~+~ \Lambda_0(i) \nonumber \\
&&~~~~~~~~~~~~~~~~~~~~+24~[~C_{i-2,i-1} +C_{i-2,i} + 2C_{i-1,i} + C_{i-1,i+1}+2C_{i,i+1}+ C_{i+1,i+2}+C_{i,i+2}~ ] \nonumber \\
&& ~~\text{where},~\nonumber \\
&&~~~~~~~~~~~~~I_{\mathcal{A}}(i)=-24~[~(C_{i-2,i-1,i} + C_{i-1,i,i-2}+C_{i,i-2,i-1}) \nonumber \\
&&~~~~~~~~~~~~~~~~~~~~~~~~~~+(C_{i-1,i,i+1}  +C_{i,i+1,i-1} +C_{i+1,i-1,i} ) \nonumber \\
&&~~~~~~~~~~~~~~~~~~~~~~~~~~+(C_{i,i+1,i+2}+C_{i+1,i+2,i}+C_{i+2,i,i+1}) ~] \label{mcalIA} \\
&&~~~~~~~~~~~~~~~~~~~~~~~~~~+4~[~(\bar{C}_{i-2,i-1}+\bar{C}_{i-2,i}) + (\bar{C}_{i-1,i-2}+\bar{C}_{i-1,i})+(\bar{C}_{i,i-2}+\bar{C}_{i,i-1}) \nonumber \\
&&~~~~~~~~~~~~~~~~~~~~~~~~~~+(\bar{C}_{i-1,i}+\bar{C}_{i-1,i+1}) + (\bar{C}_{i,i-1}+\bar{C}_{i,i+1})+(\bar{C}_{i+1,i-1}+\bar{C}_{i+1,i}) \nonumber \\
&&~~~~~~~~~~~~~~~~~~~~~~~~~~+(\bar{C}_{i,i+1}+\bar{C}_{i,i+2}) + (\bar{C}_{i+1,i}+\bar{C}_{i+1,i+2})+(\bar{C}_{i+2,i}+\bar{C}_{i+2,i+1})]
\nonumber \\
&& ~~\text{and},~\nonumber \\
&&\Lambda_0(i) = 24 \left[ (e_{i-2}e_{i-1}+e_{i-1}e_i+e_ie_{i-2}) + (e_{i-1}e_{i}+e_{i+1}e_i+e_{i-1}e_{i+1})
+(e_{i}e_{i+1}+e_{i}e_{i+2}+e_{i+1}e_{i+2})\right] \nonumber \\
&& ~~~~~~~~~~~~~~~~~~~~~~~~~~~~+57~(e_{i-2}^2+2e_{i-1}^2+2e_{i+1}^2+e_{i+2}^2) - 558~e_i^2, \nonumber
\end{eqnarray}
\noindent\rule{18cm}{0.6pt}
\begin{eqnarray}
&&12C_{i,j} = [C_{i-2,j}+2C_{i-1,j}+2C_{i+1,j}+C_{i+2,j}] + [C_{i,j-2}+2C_{i,j-1}+2C_{i,j+1}+C_{i,j+2}],~~1 \leq i < j \leq L,~|i-j|\geq 3 \nonumber
\end{eqnarray}
\noindent\rule{18cm}{0.6pt}
\begin{eqnarray}
&&240C(i,i+1) = [27C_{i-2,i+1}+15C_{i-1,i}+42C_{i-1,i+1}+42C_{i,i+2} +15C_{i+1,i+2}+27C_{i,i+3}]\nonumber \\
&&~~~~~~~~~~~~~~~~~~~~~~~+~4[K_{i-1}+2K_{i}+2K_{i+1}+K_{i+2}] +I_{\mathcal{B}_1}(i) ~+~\Lambda_1(i),~~~~~~~~~~~~~~~1 \leq i < L, \nonumber \\
&&\text{where} \nonumber \\
&&~~~~~~~~~~~~I_{\mathcal{B}_1}=-2[~\langle p_{i-1}^3(p_{i}+p_{i+1})\rangle_c + \langle p_{i}^3(p_{i-1}+p_{i+1})\rangle_c+\langle p_{i+1}^3(p_{i-1}+p_{i})\rangle_c
\nonumber \\
&&~~~~~~~~~~~~~~~~~~~~~~+~~~\langle p_{i}^3(p_{i+1}+p_{i+2})\rangle_c + \langle p_{i+1}^3(p_{i}+p_{i+2})\rangle_c+\langle p_{i+2}^3(p_{i}+p_{i+1})\rangle_c~]
\label{mcalIB1} \\
&&~~~~~~~~~~~~~~~~~~~~~~~+12[~\langle p_{i-1}^2p_{i}p_{i+1}\rangle_c + \langle p_{i-1}p_{i}^2p_{i+1}\rangle_c+\langle p_{i-1}p_{i}p_{i+1}^2\rangle_c \nonumber \\
&&~~~~~~~~~~~~~~~~~~~~~~~+~~~~\langle p_{i}^2p_{i+1}p_{i+2}\rangle_c + \langle p_{i}p_{i+1}^2p_{i+2}\rangle_c+\langle p_{i}p_{i+1}p_{i+2}^2\rangle_c~], \nonumber \\
&&\text{and} \nonumber \\
&&~~~~~~~~\Lambda_1(i) = 27e_{i-2}e_{i+1}+15e_{i-1}e_{i}+42e_{i-1}e_{i+1}+42e_{i}e_{i+2}+15e_{i+1}e_{i+2}+27e_{i}e_{i+3}] \nonumber \\
&&~~~~~~~~~~~~~~~~~~~~~~~~~~~~~~~~~~~+~12[e_{i-1}^2+2e_{i}^2+2e_{i+1}^2+e_{i+2}^2]- 240 e_{i}e_{i+1} \nonumber
\end{eqnarray}
\begin{eqnarray}
&&240C(i,i-1) = [27C_{i,i-3}+15C_{i-1,i-2}+42C_{i,i-2}+42C_{i+1,i-1}+15C_{i+1,i}+27C_{i+2,i-1}] \nonumber \\
&&~~~~~~~~~~~~~~~~~~~~~~~+~4[K_{i-2}+2K_{i-1}+2K_{i}+K_{i+1}] +I_{\mathcal{B}_{-1}}(i) ~+~\Lambda_1(i),,~~~~~~~~~~1 < i \leq L, \nonumber \\
&&\text{where},~~~~~~~~~~~~~~~~~~~~~~~~~~~~~~~~~I_{\mathcal{B}_{-1}}(i)=I_{\mathcal{B}_{1}}(i-1) \label{mcalIBm1}
\end{eqnarray}
\noindent\rule{18cm}{0.6pt}
\begin{eqnarray}
&&282C(i,i+2) = [27C_{i-2,i+2}+54C_{i-1,i+2}+42C_{i+1,i+2}+42C_{i,i+1} + 54C_{i,i+3}+27C_{i,i+4}]\nonumber \\
&&~~~~~~~~~~~~~~~~~~~~~~ + 4[K_{i}+K_{i+1}+K_{i+2}]+I_{\mathcal{B}_2}(i) +\Lambda_2(i),~~~~~~~~~~1 \leq i < L-1, \nonumber \\
&&\text{where} \nonumber \\
&&~~~~~~~~~~~~~~I_{\mathcal{B}_2}=-2[~\langle p_{i}^3(p_{i+1}+p_{i+2})\rangle_c + \langle p_{i+1}^3(p_{i}+p_{i+2})\rangle_c
+\langle p_{i+2}^3(p_{i}+p_{i+1})\rangle_c~] \label{mcalIB2} \\
&&~~~~~~~~~~~~~~~~~~~~~~+12[~\langle p_{i}^2p_{i+1}p_{i+2}\rangle_c + \langle p_{i}p_{i+1}^2p_{i+2}\rangle_c+\langle p_{i}p_{i+1}p_{i+2}^2\rangle_c~]
 +\Lambda_2(i) \nonumber \\
&&\text{where} \nonumber \\
&&~~~~~~~~~~~~~~\Lambda_2(i) = [27e_{i-2}e_{i+2}+54e_{i-1}e_{i+2}+42e_{i+1}e_{i+2}+42e_{i}e_{i+1} \nonumber \\
&&~~~~~~~~~~~~~~~~~~~~~~~~~+ 54e_{i}e_{i+3}+27e_{i}e_{i+4}] + 12[e_{i}^2+e_{i+1}^2+e_{i+2}^2]-282e_{i}e_{i+2}\nonumber \\
\end{eqnarray}
\begin{eqnarray}
&&282C(i,i-2) = [27C_{i,i-4}+54C_{i,i-3}+42C_{i,i-1}+42C_{i-1,i-2} + 54C_{i+1,i-2}+27C_{i+2,i-2}]\nonumber \\
&&~~~~~~~~~~~~~~~~~~~~~~ + 4[K_{i-2}+K_{i-1}+K_{i}]+I_{\mathcal{B}_{-2}}(i) +\Lambda_2(i),~~~~~~~~~~2 < i \leq L, \nonumber \\
&&\text{where} ,~~~~~~~~~~~~~~~~~~~~~~~~~~~~~~~~~I_{\mathcal{B}_{-2}}(i)=I_{\mathcal{B}_{2}}(i-2).  \label{mcalIBm2}
\end{eqnarray}

\noindent\rule{18cm}{0.6pt}
\begin{eqnarray}
&& 324\langle p_i^3p_{i+1}\rangle_c = 21[\langle p_{i-2}^3p_{i+1}\rangle_c
+ \langle p_{i-1}^3p_{i+1}\rangle_c+\langle p_{i}^3p_{i+1}\rangle_c] + 27[~\langle p_i^3(p_{i+1}+p_{i+2}+p_{i+3})\rangle_c~] \nonumber \\
&&~~~~~~~~~~~~~~~~~~~~~+13[\langle p_{i-1}^3(p_i+p_{i+1}) \rangle_c + \langle p_{i}^3(p_{i-1}+p_{i+1}) \rangle_c
+ \langle p_{i+1}^3(p_{i-1}+p_{i}) \rangle_c~] \nonumber \\
&&~~~~~~~~~~~~~~~~~~~~~+13[\langle p_{i}^3(p_{i+1}+p_{i+2}) \rangle_c
+ \langle p_{i+1}^3(p_{i}+p_{i+2}) \rangle_c + \langle p_{i+2}^3(p_{i}+p_{i+1}) \rangle_c~] \label{p3p} \\
&&~~~~~~~~~~~~~~~~~~~~~+9[~\langle p_{i-2}^2~(p_{i-1}+p_{i})~p_{i+1} \rangle_c
+\langle p_{i-1}^2~(p_{i-2}+p_{i})~p_{i+1} \rangle_c+\langle p_{i}^2~(p_{i-1}+p_{i-2})~p_{i+1} \rangle_c ~] \nonumber \\
&&~~~~~~~~~~~~~~~~~~~~~+3[~\langle p_{i-1}^2p_{i}p_{i+1} \rangle_c +\langle p_{i-1}p_{i}^2p_{i+1} \rangle_c +\langle p_{i-1}p_{i}p_{i+1}^2 \rangle_c~]
+3[~\langle p_{i}^2p_{i+1}p_{i+2} \rangle_c +\langle p_{i}p_{i+1}^2p_{i+2} \rangle_c +\langle p_{i}p_{i+1}p_{i+2}^2 \rangle_c~] \nonumber \\
&&~~~~~~~~~~~~~~~~~~~~~+[~K_{i-1}+2K_{i}+2K_{i+1}+K_{i+2}~] - 3[~C(i-1,i)+C(i-1,i+1)+2C(i,i+1) \nonumber \\
&&~~~~~~~~~~~~~~~~~~~~~+C(i,i+2)+C(i+1,i+2)~] - 36 \langle p_{i-2}p_{i-1}p_{i}p_{i+1}\rangle_c\nonumber \\
&&~~~~~~~~~~~~~~~~~~~~~+ 3[~e_{i-1}^2+2e_{i}^2+2e_{i+1}^2+e_{i+2}^2~]
-3[~e_{i-1}e_{i}+e_{i-1}e_{i+1}+2e_{i}e_{i+1} +e_{i}e_{i+2} +e_{i+1}e_{i+2}] \nonumber
\end{eqnarray}
\noindent\rule{18cm}{0.6pt}
\begin{eqnarray}
&&405\langle p_{i-1}p_i^2p_{i+1}\rangle_c=27[~\langle (p_{i-3}+p_{i-2}+p_{i-1})p_{i}^2p_{i+1}\rangle_c
+ \langle (p_{i+1}+p_{i+2}+p_{i+3})p_{i}^2p_{i-1}\rangle_c~] \nonumber \\
&&~~~~~~~~~~~~~~~~~~~~~~~~+ 21~[\langle p_{i-1}^2p_{i}p_{i+1}\rangle_c + \langle p_{i-1}p_{i}^2p_{i+1}\rangle_c
+ \langle p_{i-1}p_{i}p_{i+1}^2\rangle_c]+18\langle p_{i-2}p_{i-1}p_{i}p_{i+1}\rangle_c\nonumber \\
&&~~~~~~~~~~~~~~~~~~~~~~~~+9[~\langle p_{i-2}^2(p_{i-1}+p_{i})p_{i+1}\rangle_c
+ \langle p_{i-1}^2(p_{i-2}+p_{i})p_{i+1}\rangle_c +\langle p_{i}^2(p_{i-2}+p_{i-1})p_{i+1}\rangle_c~] \label{pp2p} \\
&&~~~~~~~~~~~~~~~~~~~~~~~~+9[~\langle p_{i}^2(p_{i+1}+p_{i+2})p_{i-1}\rangle_c
+ \langle p_{i+1}^2(p_{i}+p_{i+2})p_{i-1}\rangle_c +\langle p_{i+2}^2(p_{i}+p_{i+1})p_{i-1}\rangle_c~] \nonumber \\
&&~~~~~~~~~~~~~~~~~~~~~~~~+3[~\langle (p_{i-2}^3+p_{i-1}^3+p_{i}^3)p_{i+1}\rangle_c+\langle (p_{i}^3+p_{i+1}^3+p_{i+2}^3)p_{i-1}\rangle_c~]
 +18\langle p_{i-1}p_{i}p_{i+1}p_{i+2}\rangle_c  \nonumber \\
&&~~~~~~~~~~~~~~~~~~~~~~~~+[~\langle p_{i-1}^3(p_{i}+p_{i+1})\rangle_c+\langle p_{i}^3(p_{i-1}+p_{i+1})\rangle_c
+\langle p_{i+1}^3(p_{i-1}+p_{i})\rangle_c~] \nonumber \\
&&~~~~~~~~~~~~~~~~~~~~~~~~+18[~C(i-1,i)+ C(i,i+1) +C(i-1,i+1)~] -6[~K_{i-1}+K_{i}+K_{i+1}~] \nonumber \\
&&~~~~~~~~~~~~~~~~~~~~~~~~+18[e_{i-1}e_{i}+e_{i}e_{i+1}+e_{i-1}e_{i+1}-e_{i-1}e_{i-1}-e_{i}e_{i}-e_{i+1}e_{i+1}~] \nonumber
\end{eqnarray}
\noindent\rule{18cm}{0.6pt}
\begin{eqnarray}
&&132 \langle p_{i}p_{i+1}p_{i+2}p_{i+3}\rangle_c= 9[~\langle (~p_{i-2}+p_{i-1}+p_{i}~)~p_{i+1}p_{i+2}p_{i+3}\rangle_c
+ \langle p_{i}p_{i+1}p_{i+2}(~p_{i+3}+p_{i+4}+p_{i+5}~)~\rangle_c~] \nonumber \\
&&~~~~~~~~~~~~~~~+3[~\langle (~p_{i-1}p_{i}+p_{i}p_{i+1}+p_{i-1}p_{i+1}~)~p_{i+2}p_{i+3}\rangle_c
+\langle (~p_{i+2}p_{i+3}+p_{i+3}p_{i+4}+p_{i+2}p_{i+4}~)~p_{i}p_{i+1}\rangle_c~] \nonumber \\
&&~~~~~~~~~~~~~~~+3[~\langle p_{i}^2(p_{i+1}+p_{i+2})p_{i+3} \rangle_c +\langle p_{i+1}^2(p_{i}+p_{i+2})p_{i+3} \rangle_c
+\langle p_{i+2}^2(p_{i}+p_{i+1})p_{i+3} \rangle_c] \label{pppp} \\
&&~~~~~~~~~~~~~~~+3[~\langle p_{i+1}^2(p_{i+2}+p_{i+3})p_{i} \rangle_c +\langle p_{i+2}^2(p_{i+1}+p_{i+3})p_{i} \rangle_c
+\langle p_{i+3}^2(p_{i+1}+p_{i+2})p_{i} \rangle_c~] \nonumber \\
&&~~~~~~~~~~~~~~~-2[~\langle (p_{i}^3+p_{i+1}^3+p_{i+2}^3)p_{i+3} \rangle_c+\langle (p_{i+1}^3+p_{i+2}^3+p_{i+3}^3)p_{i} \rangle_c~] \nonumber
\end{eqnarray}

\section{}
\label{appendix-3}
In this appendix we justify that the correlations $\bar{C}_{i,j}=\langle p_i^3 p_j \rangle_c$, $C_{i,j,k}=\langle p_i^2 p_j p_k \rangle_c$ 
appearing in the equations (\ref{4th-mom-corr})-(\ref{C-aprox-2}) of $K_i$ and $C_{i,j}$,  do not contribute at order $\mathcal{O}(1/L)$.
To do so we first assume that these correlations
also have following scaling forms
\begin{equation}
\bar{C}_{i,j} \simeq \mathscr{U}_L\left(\frac{i}{L},\frac{j}{L}\right),~~
C_{i,j,k} \simeq \mathscr{V}_L\left(\frac{i}{L},\frac{j}{L},\frac{k}{L}\right),~~\text{and}~~
C_{i,j,k,l} \simeq \mathscr{Z}_L\left(\frac{i}{L},\frac{j}{L},\frac{k}{L},\frac{l}{L}\right), \label{oth-corr-sclng}
\end{equation}
in the $L \to \infty$ limit. Next, in the same limit we find that
the contributions from terms like $C_{i,j,k}$ and $\bar{C}_{i,j}$ appear in
eqs. (\ref{4th-mom-corr})-(\ref{corr-eq-apndx-1}) only in the
combination $\mathcal{U}_L(x)-3\mathcal{V}_L(x)$,
where $\mathcal{U}_L(x) =\mathscr{U}_L(x,y)|_{|x-y| \to 0}$ and
$\mathcal{V}_L(x) =\mathscr{V}_L(x,y,z)\big{|}_{ |x-y|\to 0}^{|x-z|\to 0}$
with $i=xL,~j=yL$ and $z=jL$. Using $\mathscr{K}(x) = 3 \mathscr{C}(x,y)|_{|x-y|\to 0}$ one can show that
$\mathcal{U}_L(x)-3\mathcal{V}_L(x)$ is of order $\mathcal{O}(\Delta T^2/L^2)$ and for that,
we look at the explicit eqs. (\ref{p3p}), (\ref{pp2p}) and (\ref{pppp})
satisfied by $C_{i,j,k}$, $\bar{C}_{i,j}$ and $C_{i,j,k,l}$ when $i,j,k,l$ are first or second
neighbors among themselves and lie inside the bulk. After careful observation and using eq. (\ref{K-eq-3C}) one finds that
these equations become independent of $C_{i,j}$ and $K_i$
in the leading order ($\sim \mathcal{O}(1/L)$ ). As a result we find
$\mathcal{U}_L(x)-3\mathcal{V}_L(x) \simeq \text{const.} \frac{\Delta T^2}{L^2}$. Similarly, if we assume
$\mathcal{U}_L(x)-3\mathcal{V}_L(x) \simeq \mathcal{O}(\Delta T^2/L^2)$ to begin with we obtain eq. (\ref{K-eq-3C}). Hence, the correlations 
$\bar{C}_{i,j}$ and $C_{i,j,k}$ in the equations of $K_i$ and $C_{i,j}$ [see eqs. (\ref{4th-mom-corr})-(\ref{corr-eq-apndx-1})], may 
contribute only to the strength of the individual inhomogeneous terms of the order $\sim \Delta T^2/L^2$. However, we have shown in eq. (\ref{mcalIeq0}) 
that their combined contribution is zero even at order $\mathcal{O}(1/L^2)$ in the continuum limit.


\end{document}